\documentclass[preprint,10pt]{elsarticle}

\usepackage{amsmath, amsthm, mathtools, amssymb, color, float, eqnarray, booktabs, makecell, soul, mdframed, multirow}
\usepackage[hyphens]{url}
\usepackage[justification=centering]{caption}
\usepackage{todonotes}
\usepackage{comment}
\usepackage{bm}
\usepackage{ulem}
\usepackage[bookmarks,colorlinks=true,urlcolor=black,linkcolor=black,citecolor=black]{hyperref}
\usepackage{cleveref}
\usepackage{subcaption}

\pdfoutput=1

\usepackage[most]{tcolorbox}

\usepackage[utf8]{inputenc}
\usepackage[english]{babel}
\usepackage{soul}

\journal{arXiv.org}

\begin{document}

\begin{frontmatter}

\title{AI-Lorenz: A physics-data-driven framework for black-box and gray-box identification of chaotic systems with symbolic regression}


\author[label1]{Mario De Florio\corref{corresponding}}
\author[label3,label4]{Ioannis G. Kevrekidis}
\author[label1,label2]{George Em Karniadakis}

\address[label1]{Division of Applied Mathematics, Brown University, Providence, RI, USA}
\address[label2]{School of Engineering, Brown University, Providence, RI, USA}
\address[label3]{Department of Chemical and Biomolecular Engineering, Johns Hopkins University, Baltimore, MD, USA}
\address[label4]{Department of Applied Mathematics and Statistics, Johns Hopkins University, Baltimore, MD, USA}

\cortext[corresponding]{Corresponding author.\\
Email address: mario\_de\_florio@brown.edu}

\address{}

\begin{abstract}
Discovering mathematical models that characterize the observed behavior of dynamical systems remains a major challenge, especially for systems in a chaotic regime. The challenge is even greater when the physics underlying such systems is not yet understood, and scientific inquiry must solely rely on empirical data. Driven by the need to fill this gap, we develop a framework that learns 
mathematical expressions modeling complex dynamical behaviors by identifying differential equations from noisy and sparse observable data. We train a small neural network to learn the dynamics of a system, its rate of change in time, and missing model terms, which are used as input for a symbolic regression algorithm to autonomously distill the explicit mathematical terms. This, in turn, enables us to predict the future evolution of the dynamical behavior. The performance of this framework is validated by recovering the right-hand sides and unknown terms of certain complex, chaotic systems such as the well-known Lorenz system, a six-dimensional hyperchaotic system, and the non-autonomous Sprott chaotic system, and comparing them with their known analytical expressions.

\end{abstract}

\begin{keyword}
Chaotic systems \sep Black-Box identification \sep Gray-Box identification \sep Random projection neural networks \sep Symbolic regression \sep Time series \sep Functional interpolation 
\end{keyword}

\end{frontmatter}


\section{Introduction}\label{sec:introduction}
Understanding the behavior of complex systems has long been a central pursuit in physics. The ability to predict the evolution of such systems is of fundamental importance for numerous fields of science, including weather forecasting, economics, and biology. However, the challenge persists when the underlying physical principles are not fully understood, making it necessary to rely solely on observable data. Many works have highlighted the significance of both black-box and gray-box models in the context of dynamical system identification \cite{rico1994continuous,garcia2022machine,yuan2023multi,sun2023chaotic} and non-autonomous dynamics with periodic forcing \cite{clemson2014discerning}. While gray-box models incorporate some knowledge of the physics of the phenomena under study, black-box models solely rely on data-driven approaches. Some of the earliest works in the direction of system identification using neural networks (NN) based methods are given by  \cite{hudson1990nonlinear,krischer1993model,kevrekidis1994global,rico1992discrete}, to obtain models representing the dynamical behavior of the chemical and fluid dynamics systems considered. 

In an early paper, González-García et al. \cite{gonzalez1998identification} proposed an NN-based methodology for the identification of distributed parameter systems. The algorithm couples a finite differences scheme with integration using the method of lines, to create the template for the NN identifier. Then, the NN is used to approximate the right-hand-side (RHS) of the partial differential equation (PDE), based on spatially distributed and temporally resolved measurements of the state variable. 
In more recent work \cite{zhu2023implementation}, the authors aim to learn the unknown dynamics from data, by using ODE-nets based on implicit numerical initial value problem (IVP) solvers, thus
identifying a so-called ``inverse modified differential equation"
of the system. This framework can be easily adapted to solve gray-box identification, by incorporating partially known physical terms. An interesting application of NN for directly learning evolution equations (black-box), parameters discovery (white-box), and learning partially unknown kinetics terms (gray-box) is given by Cui et al. \cite{cui2023data}, which was applied to Chinese hamster ovary cell bioreactors by using process data. 
Recently, new deep NNs frameworks have been developed to forecast complex dynamics for gray-box models \cite{yin2021augmenting,malani2023some}, and applied to real-world scenarios such as building energy simulations \cite{li2021grey,thilker2021non}.

Inspired by recent advancements in deep learning, neural ordinary differential equations (neural ODEs) \cite{chen2018neural} have emerged as a promising approach to modeling complex temporal dynamics. By integrating NNs with ODEs, these models can effectively capture intricate system behaviors and their temporal evolution, allowing for more accurate predictions and insights \cite{linot2023stabilized}.
In the past few years, developments have arisen by merging neural ODEs with SINDy to discover governing equations from data \cite{fronk2023interpretable,goyal2022discovery,lee2022structure} with symbolic regression (SR). SINDy (Sparse Identification of Nonlinear Dynamics), introduced by Brunton et al. \cite{brunton2016discovering}, is a novel approach for addressing the dynamical system discovery challenge, utilizing the principles of sparse regression \cite{tibshirani1996regression} and compressed sensing \cite{donoho2006compressed}. Their methodology builds on the description of many physical systems as a small set of pertinent terms governing their dynamics, thus resulting in sparse formulations within a high-dimensional nonlinear function space. SINDy has demonstrated its ability to discover nonlinear dynamical systems from data \cite{champion2019data, proctor2014exploiting,bakarji2023discovering,wei2022sparse}, as long as the terms of the sought equations lie in the span of the candidate library.

Udrescu et al. \cite{udrescu2020ai} introduced a framework for extracting governing equations from data. Their work utilizes SR to identify a symbolic expression that accurately represents an unknown function based on a given dataset. This new recursive multidimensional symbolic regression algorithm, known as \textit{AI-Feynman}, merges NN methodologies with physics-inspired strategies. It has been tested through the discovery of 100 closed-form equations from the \textit{Feynman Lectures on Physics}, and compared with existing publicly available software. Despite the groundbreaking potential of this research, it is limited to discovering equations involving elementary functions, not including those involving derivatives and integrals, which are widely present in real-world phenomena. Certainly, improving this aspect would be a valuable addition to the framework. 

Overall, the advancement and recent improvements of SR algorithms are showing interesting prospects for future breakthroughs in the domain of physics and beyond. In the \textit{AI-Descartes} framework \cite{cornelio2023combining}, the authors derive meaningful mathematical models from a fusion of axiomatic knowledge and empirical data by integrating logical reasoning with SR \cite{marra2019constraint,scott2020lgml,ashok2021logic}. The unique aspect of this method lies in its endeavor to generate models that align with overarching logical axioms. In a more recent work \cite{daryakenaria2023ai}, \textit{AI-Aristotle} demonstrated superior performance in the field of gray-box identification of systems of ODEs. In particular, the authors proposed two NN-based algorithms for parameters and unknown terms discovery (namely X-TFC \cite{schiassi2021extreme,de2022physics} and PINNs \cite{raissi2019physics}), applied to two systems biology problems, and explicitly discovered the mathematical terms of the differential equations using two SR methods, PySR \cite{cranmer2023interpretable} and gplearn \cite{stephens2015gplearn} for cross-validation. An interesting recent work by Boddupalli et al. \cite{boddupalli2023symbolic} proposes an algorithm to identify closed-form models of dynamical systems from time-series data. The work uses a deep neural network able to generate a symbolic expression for the governing equations, by searching for a space of mathematical expressions to find the model that optimizes the parameters for a fixed dictionary.
    
In the current paper, we propose an alternative framework to discover black-box and gray-box models of dynamical systems, in a purely data-driven scenario. This framework introduces Black-Box X-TFC (BBX-TFC) and Gray-Box X-TFC (GBX-TFC) as efficient NN-based methods for dynamics data-fitting, and right-hand-side (RHS) and missing terms identification, which will feed the PySR algorithm for distilling the governing equations of the systems in consideration. We test our method with autonomous and non-autonomous chaotic and hyperchaotic systems because of their high sensitivity to small parameter and initial condition perturbations. Nonetheless, we expect that the application of the method to any other time-series dataset will be straightforward. 

The paper is organized as follows. In Section \ref{sec:methodology}, we introduce the two main methods, with a particular focus on the X-TFC algorithm applied to the Lorenz system. We present the results of our test cases in Section \ref{sec:results} for the Lorenz system, 6D hyperchaotic systems, and non-autonomous Sprott system, demonstrating the performance of X-TFC for dynamics regression, and RHS and missing term extraction, and the performance of PySR for mathematical expressions distillation. Finally, we summarize our findings in Section \ref{sec:conclusions}.

\section{Methodology}\label{sec:methodology}

\begin{figure}[h!]
    \centering
    \includegraphics[width=\linewidth]{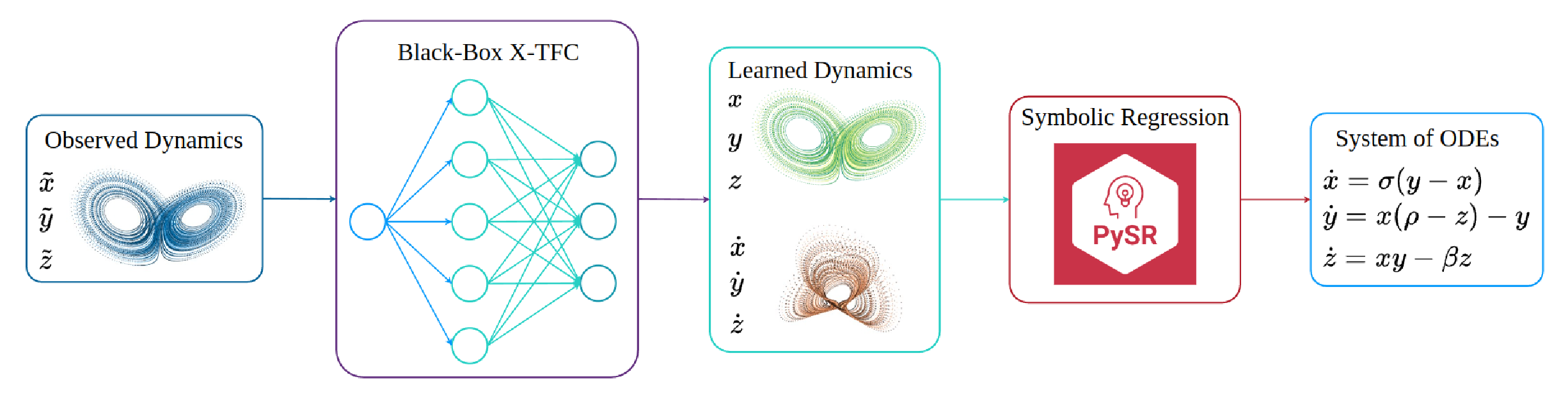}
    \caption{AI-Lorenz framework: the observed dynamics are used to train the neural network in the black-box X-TFC framework, which learns the dynamics and their rates of change in time. The latter are used as input in the symbolic regression algorithm (PySR), which autonomously distills the explicit mathematical terms that build the system of differential equations.}
    \label{fig:ai_lorenz_scheme}
\end{figure}

In this section, we introduce the two core components of AI-Lorenz: X-TFC and PySR. BBX-TFC is the adaptation of the X-TFC framework \cite{schiassi2021extreme} in a physics-agnostic fashion for learning the dynamics of complex systems, and their rate of change in time, purely from data, while GBX-TFC is a physics-data-driven framework that can also use partial knowledge of the physics in combination with the observed data. 
The output from BBX-TFC and GBX-TFC are used as input to PySR \cite{cranmer2023interpretable}, an open-source tool for SR able to find interpretable symbolic expressions that optimize some objectives. In our case, it will find the exact mathematical expressions of the RHS of the differential equations of the dynamical systems. A schematic of the overall operation of the AI-Lorenz framework for black-box identification (available at \href{https://github.com/mariodeflorio/AI-Lorenz}{https://github.com/mariodeflorio/AI-Lorenz}) is shown in Figure \ref{fig:ai_lorenz_scheme}. Let us now examine in detail how the two main components of this machine work. 

\subsection{Black-Box X-TFC}

The Theory of Functional Connections (TFC) \cite{mortari2017theory} provides the so-called constrained expression (CE) to approximate the solution of the differential equation, in a form depending on the problem constraints \cite{leake2020multivariate,de2021theory,mai2022theory}. For an ODE of the form $\frac{dy}{dt}=f(y,t)$ subject to initial condition $y(0)=y_0$, the unknown solution of the ODE is represented with the following CE \cite{mortari2017least}:
\begin{equation}\label{eq:ce}
    y(t,\boldsymbol{\beta}) = g(t,\boldsymbol{\beta}) - g(0,\boldsymbol{\beta}) + y_0,
\end{equation}
where $g(t,\boldsymbol{\beta})$ is a free-chosen function, selected by the user. It is trivial to note that whatever function is chosen, the CE will always satisfy the initial condition analytically. According to the X-TFC framework \cite{schiassi2021extreme}, we choose $F(t)$ to be a single-layer random projection neural network, in which input weights and biases are initially randomly selected, before the training process. Mathematically, this NN can be expressed as
\begin{equation}\label{eq:singleNN}
g(t,\boldsymbol{\beta}) = \sum_{j=1}^{L} \beta_j\sigma \left(w_jt + b_j \right)= \begin{bmatrix}
\sigma(w_1 t + b_1) \\ \vdots \\  \sigma(w_L t + b_L) 
\end{bmatrix}^T  \boldsymbol{\beta} = \boldsymbol{\sigma}^T \boldsymbol{\beta},
\end{equation}
where $j=1,...,L$ represents the index of the \textit{j-th} neuron of the NN hidden layer, $w_j$ and $b_j$ represent its assigned weight and bias, respectively, and $\beta_j$ is its output weight. The neurons are defined by the activation function $\sigma(wx + b)$, which is also selected by the user. For this work, the chosen activation function is $tanh$, thus 
\begin{equation}
    \sigma(w_j x + b_j) = \frac{e^{2(w_j x + b_j)-1}}{e^{2(w_j x + b_j)+1}}.
\end{equation}
Thus, the CE and its derivative can be respectively rewritten as
\begin{subequations}
\begin{align}
    y(t,\boldsymbol{\beta}) &= \bigl[ \boldsymbol{\sigma}(t) - \boldsymbol{\sigma}(0) \bigr] ^T \boldsymbol{\beta} + y_0 \\
    \dot{y}(t,\boldsymbol{\beta}) &= \dot{\boldsymbol{\sigma}}(t) ^T \boldsymbol{\beta}, 
\end{align}
\end{subequations}
where $\boldsymbol{\sigma}(t)$ and $\boldsymbol{\sigma}(0)$, from now on, will be referred as $\boldsymbol{\sigma}$ and $\boldsymbol{\sigma}_0$, respectively, for the sake of simplicity.\\

\begin{figure}[h!]
    \centering
    \includegraphics[width=0.8\linewidth]{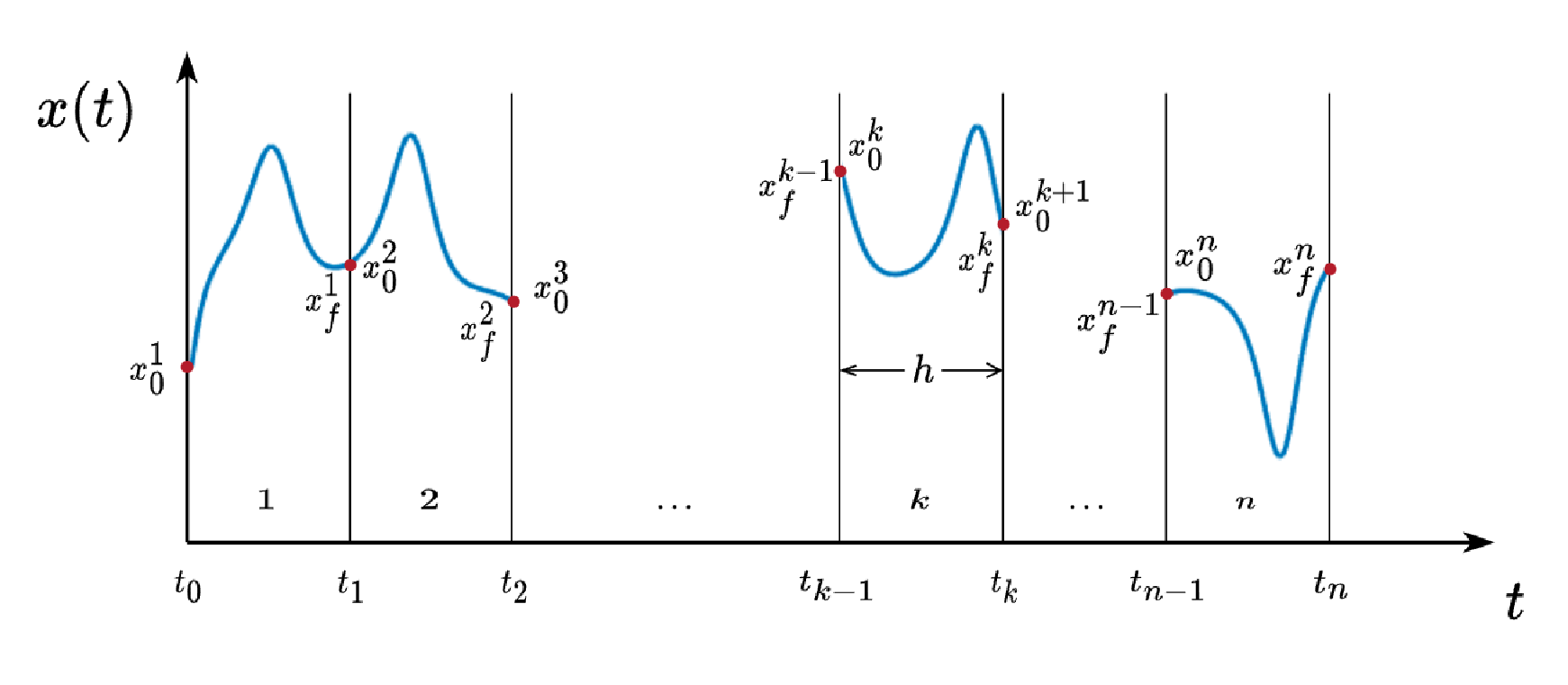}
    \caption{X-TFC time domain decomposition scheme. The domain is divided into equispaced subdomains with a time-step of length $h$. Continuity conditions are imposed at the interface of the subdomains.}
    \label{fig:domain_decomp}
\end{figure}

Since the problems tackled in this work are typically stiff, chaotic for certain parameters' values, and with relatively large integration time, 
a time-domain decomposition is performed \cite{de2022physics,schiassi2022physics} by dividing the domain into $n$ subintervals of equal length $h = t_k - t_{k-1}$, for $k=1,...,n$. This leads to the composition of several consecutive initial value problems (IVPs) of the form $\frac{dy{^{(k)}}}{dt}=f(y{^{(k)}},t)$ subject to initial condition $y{^{(k)}}(0)=y_0{^{(k)}}$. Thus, the X-TFC framework will be applied consecutively for each subdomain, imposing a continuity condition at the subdomain interfaces 
\begin{equation}
    y_0^{(k)} = y_f^{(k-1)}
\end{equation}
as shown in Figure \ref{fig:domain_decomp}.

\begin{figure}[h!]
    \centering
    \includegraphics[width=\linewidth]{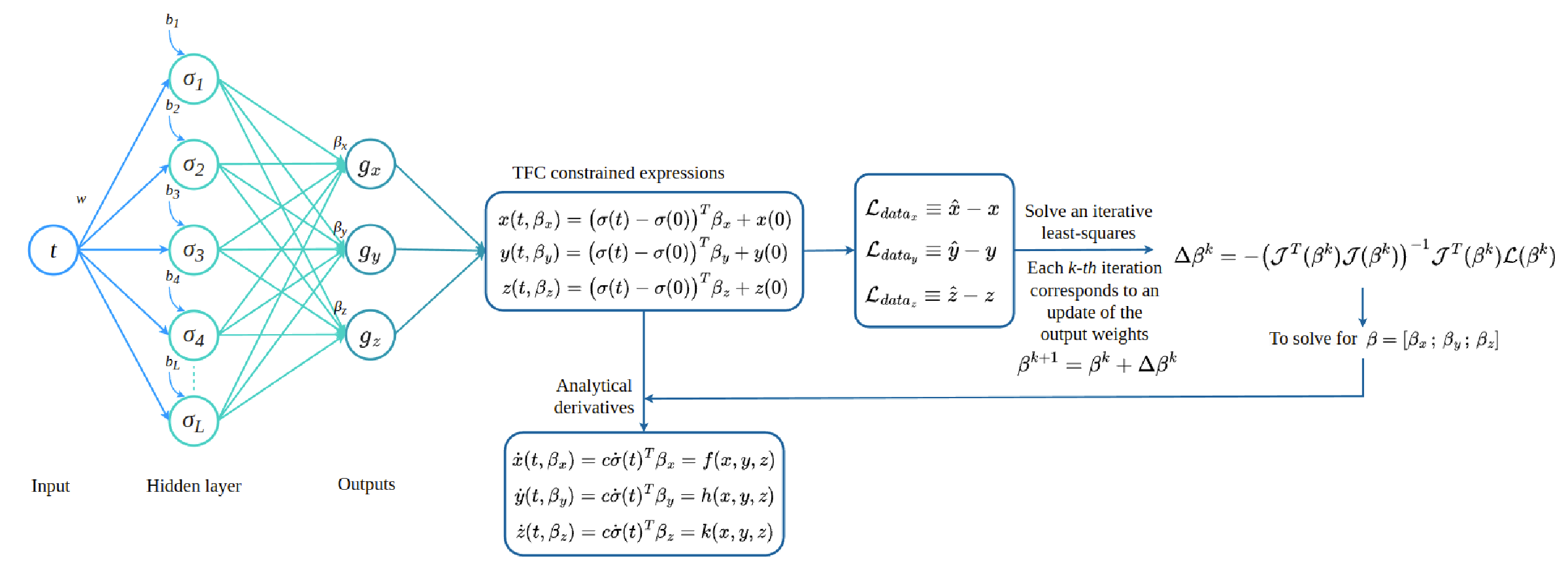}
    \caption{Schematic of the X-TFC algorithm for performing black-box identification of a 3D chaotic system. Input weights and biases are randomly selected. The last step solves iteratively a least-squares, thus no back-propagation is involved in the training, allowing fast computational times.}
    \label{fig:schematic_xtfc}
\end{figure}

Now that the CE for an IVP is defined, we show how to perform X-TFC black-box identification using the Lorenz system of \eqref{eq:lorenz} as an example, at each subdomain. Straightforwardly, the same framework can be applied to any nonlinear dynamical system, autonomous or non-autonomous, modeled by ODEs, for both black-box and gray-box identification. The black-box model to be discovered can be represented by the following system
\begin{equation}\label{eq:lorenz_black}
    \begin{cases}
\dot{x} = f(x,y,z) \\
\dot{y} = h(x,y,z)  \\
\dot{z} = k(x,y,z)
    \end{cases} 
\end{equation}
in which the right-hand-side functions $f(x,y,z),h(x,y,z),$ and $k(x,y,z)$ are totally unknown. The first step is to build the CEs for each state variable under consideration. For this example, we have
\begin{subequations}\label{eq:lorenz_ce}
\begin{align}
    x = \bigl( \boldsymbol{\sigma} - \boldsymbol{\sigma}_0 \bigr) ^T \boldsymbol{\beta}_{x} + x(0)    \\
    y = \bigl( \boldsymbol{\sigma} - \boldsymbol{\sigma}_0 \bigr) ^T \boldsymbol{\beta}_{y} + y(0)  \\
    z = \bigl( \boldsymbol{\sigma} - \boldsymbol{\sigma}_0 \bigr) ^T \boldsymbol{\beta}_{z} + z(0),
\end{align}
\end{subequations}
and their derivatives
\begin{subequations}\label{eq:lorenz_ce_der}
\begin{align}
    \dot{x} = c \dot{\boldsymbol{\sigma}} ^T \boldsymbol{\beta}_{x}   \\
    \dot{y} = c \dot{\boldsymbol{\sigma}} ^T  \boldsymbol{\beta}_{y}  \\
    \dot{z} = c \dot{\boldsymbol{\sigma}} ^T \boldsymbol{\beta}_{z}.  
\end{align}
\end{subequations}
where $c$ is the mapping coefficient to map the time domain $t$ into the activation function domain $[-1,1]$. The loss functions we want to minimize are the differences between the observed dynamics $(\tilde{x},\tilde{y},\tilde{z})$ and their CEs approximations:
\begin{subequations}\label{eq:loss_indirect}
\begin{align}
    \mathcal{L}_{data_{x}} & \equiv \tilde{x} - x    \\
    \mathcal{L}_{data_{y}} & \equiv \tilde{y} - y  \\
    \mathcal{L}_{data_{z}} & \equiv  \tilde{z} - z.
\end{align}
\end{subequations}
The next step is to build a Jacobian matrix, by deriving the loss functions with respect to all the unknown terms (output weights $\boldsymbol{\beta}$), which will simply be the following diagonal matrix
\begin{equation}
    \boldsymbol{\mathcal{J}} = \begin{bmatrix}
        (\boldsymbol{\sigma}_0 - \boldsymbol{\sigma}) & \textbf{0} & \textbf{0}  \\
        \textbf{0} & (\boldsymbol{\sigma}_0 - \boldsymbol{\sigma}) & \textbf{0}   \\
        \textbf{0} & \textbf{0} & (\boldsymbol{\sigma}_0 - \boldsymbol{\sigma})
    \end{bmatrix} 
\end{equation}
Now, the unknown vector $\boldsymbol{\beta} = [\boldsymbol{\beta}_x; \boldsymbol{\beta}_y; \boldsymbol{\beta}_z]$ is computed by iteratively solving the linear system $\boldsymbol{\mathcal{J}} \Delta \boldsymbol{\beta}^k = \boldsymbol{\mathcal{L}}$, with $\boldsymbol{\mathcal{L}} = [\mathcal{L}_{data_{x}} ; \mathcal{L}_{data_{y}} ; \mathcal{L}_{data_{z}}]$, until preselected tolerance condition for the loss $\boldsymbol{\mathcal{L}}$ or maximum number of iterations are reached. Each \textit{k-th} iteration corresponds to an update of the output weights $\boldsymbol{\beta}^{k+1} = \boldsymbol{\beta}^k + \Delta \boldsymbol{\beta}^k$, where 
\begin{equation}
    \Delta \boldsymbol{\beta}^k = -\left[ \boldsymbol{\mathcal{J}}^T(\boldsymbol{\beta}^k)\boldsymbol{\mathcal{J}}(\boldsymbol{\beta}^k)  \right]^{-1} \boldsymbol{\mathcal{J}}^T(\boldsymbol{\beta}^k)\boldsymbol{\mathcal{L}}(\boldsymbol{\beta}^k) .
\end{equation}
Finally, the NN output weights 
%
are now trained to minimize the loss functions. Thus, by substituting them into the CEs and CEs derivatives of eqs. \eqref{eq:lorenz_ce} and \eqref{eq:lorenz_ce_der}, we obtain the learned dynamics $(x,y,z)$ and their variations in time $(\dot{x},\dot{y},\dot{z})$, or, in other words, the black-box model $f(t),h(t),$ and $k(t)$. A representative schematic of the BBX-TFC algorithm is shown in Figure \ref{fig:schematic_xtfc}, displaying its main steps.

The framework presented above embodies an indirect method to retrieve the RHS of the differential equations since what it actually computes is their rate of change (left-hand side), which is equal to the RHS. Another direction can be taken by directly learning the unknown functions $f(t),h(t),$ and $k(t)$, representing the RHS. The difference is in the set of loss functions that need to be minimized in the training process. Thus, we need to add the residuals of the differential equations of \eqref{eq:lorenz_black} to the loss system of \eqref{eq:loss_indirect}, such as
\begin{subequations}\label{eq:loss_direct}
\begin{align}
    \mathcal L_x & \equiv f(x,y,z) - \dot x \\
    \mathcal L_y & \equiv h(x,y,z) - \dot y \\
    \mathcal L_z & \equiv k(x,y,z) - \dot z \\
    \mathcal{L}_{data_{x}} & \equiv \tilde{x} - x    \\
    \mathcal{L}_{data_{y}} & \equiv \tilde{y} - y  \\
    \mathcal{L}_{data_{z}} & \equiv  \tilde{z} - z,
\end{align}
\end{subequations}
where the functions $f(x,y,z),h(x,y,z),$ and $k(x,y,z)$ are approximated with a small neural network, as follows
\begin{subequations}\label{eq:functions}
\begin{align}
    f(x,y,z) =  \boldsymbol{\sigma}^T \boldsymbol{\beta}_{f}   \\
    h(x,y,z) =  \boldsymbol{\sigma}^T \boldsymbol{\beta}_{h}   \\
    k(x,y,z) =  \boldsymbol{\sigma}^T \boldsymbol{\beta}_{k}.   
\end{align}
\end{subequations}
The next step is to build the Jacobian matrix, by deriving the loss functions with respect to all the unknown terms (output weights $\boldsymbol{\beta}$), which is the following matrix
\begin{equation}
    \boldsymbol{\mathcal{J}} = \begin{bmatrix}
    - c \dot{\boldsymbol{\sigma}}  & \textbf{0} &  \textbf{0} & \quad \boldsymbol{\sigma} \quad & \quad \textbf{0} \quad &  \quad  \textbf{0} \quad  \\
    \textbf{0} & - c \dot{\boldsymbol{\sigma}}  &  \textbf{0} & \quad \textbf{0} \quad & \ \quad \boldsymbol{\sigma}  \quad &  \quad \textbf{0} \quad  \\
    \textbf{0} &  \textbf{0}  &  - c \dot{\boldsymbol{\sigma}}  &  \quad  \textbf{0}  \quad  & \quad \textbf{0} \quad &  \quad  \boldsymbol{\sigma}  \quad \\
        (\boldsymbol{\sigma}_0 - \boldsymbol{\sigma}) &  \textbf{0}  &  \textbf{0}  & \quad \textbf{0} \quad & \quad \textbf{0} \quad &  \quad \textbf{0}   \quad \\
        \textbf{0} & (\boldsymbol{\sigma}_0 - \boldsymbol{\sigma}) & \textbf{0}  & \quad \textbf{0} \quad & \quad \textbf{0} \quad &  \quad \textbf{0}  \quad  \\
        \textbf{0} &  \textbf{0}  & (\boldsymbol{\sigma}_0 - \boldsymbol{\sigma}) & \quad \textbf{0} \quad & \quad \textbf{0} \quad & \quad  \textbf{0}  \quad
    \end{bmatrix} .
\end{equation}

\begin{figure}[h!]
    \centering
    \includegraphics[width=\linewidth]{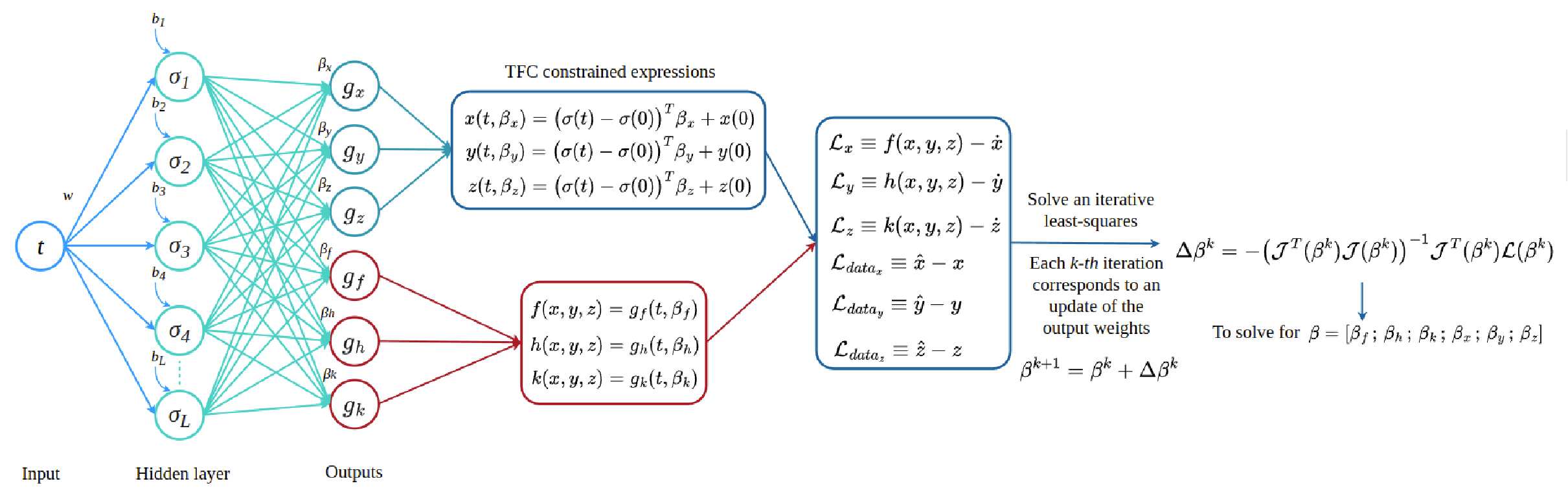}
    \caption{Schematic of the X-TFC algorithm for performing direct black-box identification of a 3D chaotic system. Input weights and biases are randomly selected. The last step solves iteratively a least-squares, thus no back-propagation is involved in the training, allowing fast computational times.}
    \label{fig:schematic_xtfc_direct}
\end{figure}

Now, the unknown vector $\boldsymbol{\beta} = [\boldsymbol{\beta}_{f} ; \boldsymbol{\beta}_{h} ; \boldsymbol{\beta}_{k} ; \boldsymbol{\beta}_x ; \boldsymbol{\beta}_y ; \boldsymbol{\beta}_z]$ is computed by iteratively solving the linear system  
$\boldsymbol{\mathcal{J}} \Delta \boldsymbol{\beta}^k = \boldsymbol{\mathcal{L}}$, with $\boldsymbol{\mathcal{L}} = [\mathcal{L}_{x} ; \mathcal{L}_{y} ; \mathcal{L}_{z} ; \mathcal{L}_{data_{x}} ; \mathcal{L}_{data_{y}} ; \mathcal{L}_{data_{z}}]$. Each \textit{k-th} iteration corresponds to an update of the output weights $\boldsymbol{\beta}^{k+1} = \boldsymbol{\beta}^k + \Delta \boldsymbol{\beta}^k$, where 
\begin{equation}
    \Delta \boldsymbol{\beta}^k = -\left[ \boldsymbol{\mathcal{J}}^T(\boldsymbol{\beta}^k)\boldsymbol{\mathcal{J}}(\boldsymbol{\beta}^k)  \right]^{-1} \boldsymbol{\mathcal{J}}^T(\boldsymbol{\beta}^k)\boldsymbol{\mathcal{L}}(\boldsymbol{\beta}^k).
\end{equation}
Finally, the NN output weights are trained to minimize the loss functions. Thus, by substituting them into the CEs, CEs derivatives, and unknown functions of eqs. \eqref{eq:lorenz_ce}, \eqref{eq:lorenz_ce_der}, and \eqref{eq:functions}, we obtain the learned dynamics $(x,y,z)$, their variations in time $(\dot{x},\dot{y},\dot{z})$, and directly the black-box model $f(x,y,z),h(x,y,z),$ and $k(x,y,z)$. A representative schematic of the direct BBX-TFC algorithm is shown in Figure \ref{fig:schematic_xtfc_direct}, displaying its main steps.
We prefer to choose the indirect method shown initially, because of its simplicity in implementation, and for its better performance in terms of accuracy and computational times, as proved in Section \ref{sec:bbxtfc_performance}.

\subsection{Gray-Box X-TFC}

\begin{figure}[h!]
    \centering
    \includegraphics[width=\linewidth]{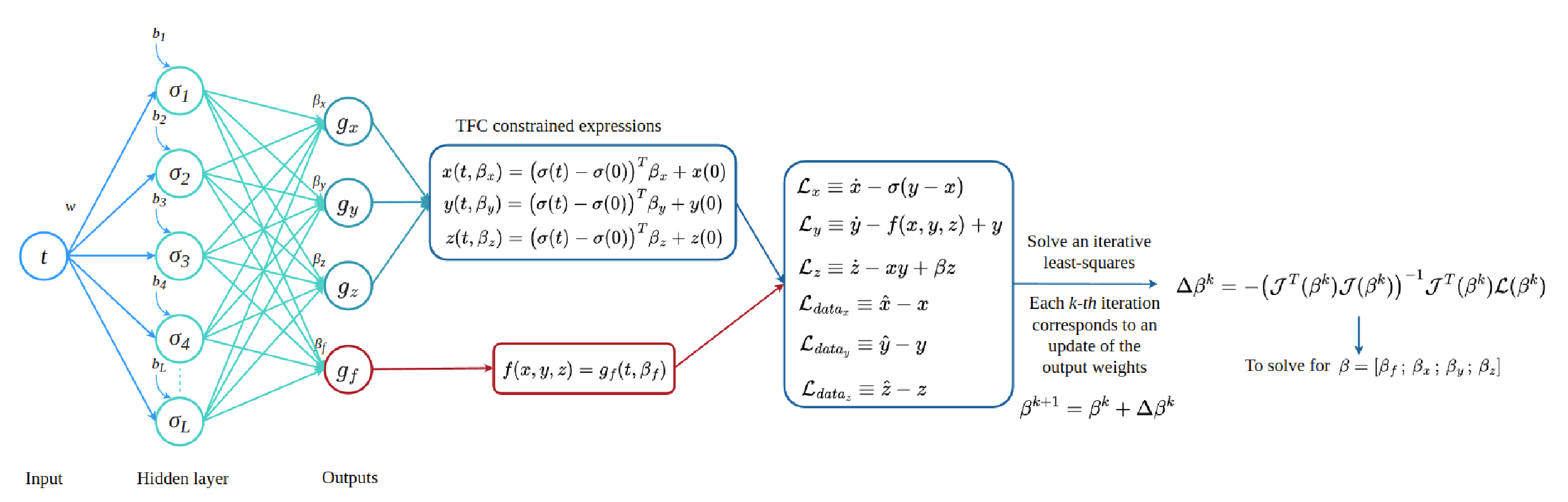}
    \caption{Schematic of the X-TFC algorithm for performing gray-box identification of a 3D chaotic system with a missing term. Input weights and biases are randomly selected. The last step solves iteratively a least-squares, thus no back-propagation is involved in the training, allowing fast computational times.}
    \label{fig:schematic_xtfc_gray}
\end{figure}

Similarly, we can perform gray-box identification when partial knowledge of the physical system terms is available, as well as observable data, within the Gray-Box X-TFC (GBX-TFC) framework. For example, let us consider the Lorenz system, in which the term $x(\rho - z)$ in the second differential equation is unknown, and the observations of the three states are available. If we name this unknown term as $f(x,y,z)$, the system of ODEs can be written as
\begin{equation}\label{eq:gray_lorenz}
    \begin{cases}
\dot{x} = \sigma (y - x) \\
\dot{y} = f(x,y,z) - y  \\
\dot{z} = xy - \beta z
    \end{cases} 
\end{equation}
One should not confuse the Lorenz parameters $\sigma$ and $\beta$ with the notations for the activation function $\boldsymbol{\sigma}$ and the output weights vector $\boldsymbol{\beta}$. \\
We can use GBX-TFC to learn the $f(x,y,z)$ function by minimizing the following set of loss functions
\begin{subequations}
\begin{align}
    \mathcal L_x & \equiv \dot x(t,\boldsymbol{\beta}_x) - \sigma\bigl(y(t,\boldsymbol{\beta}_y) - x(t,\boldsymbol{\beta}_x)\bigr) \\
    \mathcal L_y & \equiv  \dot y(t,\boldsymbol{\beta}_y) - f(t,\boldsymbol{\beta}_f) + y(t,\boldsymbol{\beta}_y) \\
    \mathcal L_z & \equiv  \dot z(t,\boldsymbol{\beta}_z) - x(t,\boldsymbol{\beta}_x) y(t,\boldsymbol{\beta}_y) + \beta z(t,\boldsymbol{\beta}_z) \\
    \mathcal{L}_{data_{x}} & \equiv \tilde{x} - x(t,\boldsymbol{\beta}_x)    \\
    \mathcal{L}_{data_{y}} & \equiv \tilde{y} - y(t,\boldsymbol{\beta}_y)  \\
    \mathcal{L}_{data_{z}} & \equiv  \tilde{z} - z(t,\boldsymbol{\beta}_z),
\end{align}
\end{subequations}
where the function $f(x,y,z)$ is approximated with a small single-layer neural network, as
\begin{equation}
    f(x,y,z) =  \boldsymbol{\sigma}^T \boldsymbol{\beta}_{f}  
\end{equation}
It follows that the Jacobian matrix, containing the derivatives of the loss functions with respect to the unknown output weights, has the following form 
\begin{equation}
    \boldsymbol{\mathcal{J}} = \begin{bmatrix}
     c \dot{\boldsymbol{\sigma}}  + \sigma(\boldsymbol{\sigma} - \boldsymbol{\sigma}_0)  & \textbf{0} &  \textbf{0} &  \quad \textbf{0} \quad  \\
    \textbf{0} &  c \dot{\boldsymbol{\sigma}} + \boldsymbol{\sigma} - \boldsymbol{\sigma}_0  &  \textbf{0} &  \ \quad -\boldsymbol{\sigma}  \quad   \\
    y(\boldsymbol{\sigma}_0 - \boldsymbol{\sigma}) &  x(\boldsymbol{\sigma}_0 - \boldsymbol{\sigma})  &   c \dot{\boldsymbol{\sigma}} + \beta (\boldsymbol{\sigma} - \boldsymbol{\sigma}_0) &  \quad  \textbf{0}  \quad  \\
        (\boldsymbol{\sigma}_0 - \boldsymbol{\sigma}) &  \textbf{0}  &  \textbf{0}  & \quad \textbf{0} \quad \\
        \textbf{0} & (\boldsymbol{\sigma}_0 - \boldsymbol{\sigma}) & \textbf{0}  & \quad \textbf{0} \quad  \\
        \textbf{0} &  \textbf{0}  & (\boldsymbol{\sigma}_0 - \boldsymbol{\sigma}) & \quad \textbf{0} \quad 
    \end{bmatrix} .
\end{equation}
From here, the procedure to compute the output weights $\boldsymbol{\beta}$ and then retrieve the dynamics and the missing term is equivalent to the above-mentioned procedure for the black-box model. A representative schematic of the GBX-TFC algorithm is shown in Figure \ref{fig:schematic_xtfc_gray}, displaying its main steps. The interested reader can find more details about BGX-TFC combined with PySR to solve gray-box models in Ref. \cite{daryakenaria2023ai}.

\subsection{PySR for Symbolic Regression}

SR is a machine learning tool in the field of data-driven modeling, allowing us to discover concise, closed-form mathematical equations that model the underlying dynamics of complex systems. Within the AI-Lorenz framework, PySR \cite{cranmer2023interpretable} serves as a fundamental engine for this purpose, since it enables the discovery of non-linear interactions that might otherwise remain hidden.

This algorithm initially generates a pool of candidate equations, represented as mathematical expressions involving fundamental operations (e.g., addition, subtraction, multiplication, division), elementary functions (such as sine, cosine, and exponentials), and variables. Subsequently, each algebraic or transcendental candidate equation undergoes evaluation against the provided dataset, with its performance assessed by a fitness function quantifying the equation's matching with the data. Typically, this fitness function calculates metrics like the mean squared error (MSE) or other similar indicators. A genetic algorithm \cite{koza1994genetic} enters the scene to select the most promising candidate equations for the subsequent generation. Equations demonstrating a better fit to the data have a higher probability of being chosen, while less fitting equations may be removed. Genetic operations like crossover (the merging of segments from two equations) and mutation (introducing small alterations to an equation) are applied to the selected equations, generating a new batch of candidate equations. This iterative process continues across multiple generations, steadily improving the equations' fitness until a termination condition, such as reaching a predefined maximum number of generations or achieving a specified fitness threshold, is met. The algorithm's in-depth details can be found in \cite{cranmer2023interpretable}.

\section{Results}\label{sec:results}

In this section, we report results for three benchmark problems: the 3D chaotic Lorenz system, a 6D hyperchaotic system, and the non-autonomous 3D Sprott attractor. In particular, we use the Lorenz system as a test case to show the performance of BBX-TFC for learning the dynamics from the trajectory data (different datasets) and its rate of change in time, representing the RHS of the system of ODEs. We compare BBX-TFC accuracy and computational time with Gaussian Process Regression (GPR), also used to feed the SR algorithm. We also compare the predicted trajectories with the discovered differential equations. Subsequently, we test the robustness of our framework by corrupting our data with certain noise levels. For the last two test cases, we show the performance of our framework to discover the equations for a higher dimensional hyperchaotic system and a non-autonomous system. The other simulations (BBX-TFC accuracy for different datasets, GPR comparison, and behavior with noise) are omitted because of their similarity in performance with the Lorenz system case.

\subsection{Lorenz System}

The first benchmark dynamical system we use for testing our framework is the well-known three-dimensional Lorenz system \cite{lorenz1963deterministic}, arising in atmospheric modeling. This system is given by the following set of nonlinear ODEs:
\begin{equation}\label{eq:lorenz}
    \begin{cases}
\dot{x} = \sigma (y - x) \\
\dot{y} = x(\rho - z) - y  \\
\dot{z} = xy - \beta z
    \end{cases} 
    \qquad \qquad \text{s.t.} \qquad \qquad 
    \begin{cases}
        x(0) = -8 \\
        y(0) = 7  \\
        z(0) = 27
    \end{cases}
\end{equation}
where the parameters are selected as $\sigma=10$, $\beta=2.6667$, and $\rho = 27$, for which the system exhibits chaotic behavior, giving rise to the strange attractor exhibiting its popular butterfly shape. The data used for training the NN has been generated with 4\textit{-th} order Runge-Kutta method for a time domain $[0,10]$, with a time step from $\Delta t = 0.05$ to $\Delta t = 0.0002$ (from 20 Hz to 5000 Hz), producing a trajectory that is attracted to the Lorenz attractor. Now, we can move forward to the discovery of differential equations in a purely data-driven fashion by using the Black-Box X-TFC. Subsequently, we briefly show the performance of the missing term discovery by using the physics-data-driven Gray-Box X-TFC algorithm.


\subsubsection{Black-Box X-TFC performance}\label{sec:bbxtfc_performance}
\begin{figure}[h!]
    \centering
    \includegraphics[width=\linewidth, trim=120 0 120 0, clip]{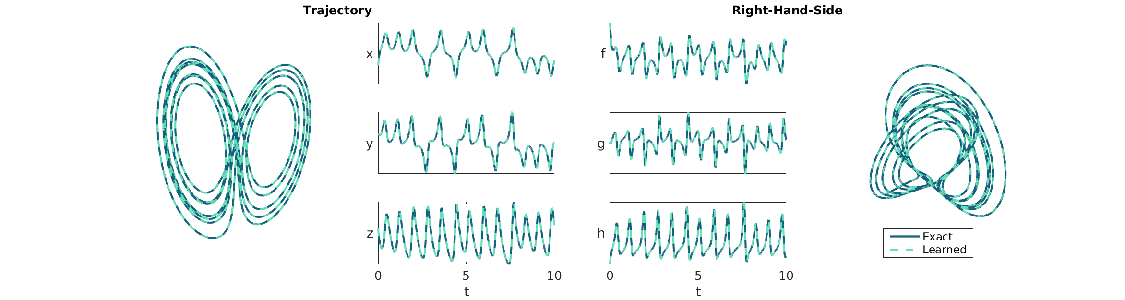}
    \caption{Lorenz system: Black-Box X-TFC learned dynamics (left) and right-hand-side (right) vs. exact solutions.}
    \label{fig:bbxtfc_lorenz}
\end{figure}

The performance of BBX-TFC is evaluated via the Mean Absolute Error (MAE) computed as:
\begin{equation*}
    MAE = \frac{\sum_{i=1}^N |\hat{h}_i(t) - h_i(t)|}{N}.
\end{equation*}
Our results are reported in Tables \ref{tab:BBXTFC_lorenz_10_10_0.1},\ref{tab:BBXTFC_lorenz_10_10_0.001}, \ref{tab:BBXTFC_lorenz_20_20_0.1}, and qualitative plots are shown in Figure \ref{fig:bbxtfc_lorenz}. The different numbers of data points generated with the Runge-Kutta method depend on the time-step length selected. The Tables report the MAE obtained with different hyperparameters of BBX-TFC used for the training process. In Table \ref{tab:BBXTFC_lorenz_10_10_0.1}, the results are obtained using 10 neurons, 10 collocation points per each sub-domain, and sub-domains of length 0.1. We see, as expected, the decrease of MAE with the increase of the number of data points, with errors in the order of 1e-12 to 1e-15 for the dynamics regression ($x,y,z$), and 1e-01 to 1e-05 for the RHS, indirectly extracted as rates of change of the dynamics. We can further reduce the MAE by reducing the length of the time-step to 0.001, while keeping the same number of neurons and collocation points, obtaining the results reported in Table \ref{tab:BBXTFC_lorenz_10_10_0.001}. We drastically reduced the errors for the learned dynamics, and slightly reduced the errors for their rates of change $(\dot x, \dot y, \dot z)$. Further decreasing of the time-step length will lead to a slight decrease of MAE for the dynamics, but no improvement for the right-hand side. Eventually, we found a better fine-tuning in the parameters, such as 20 neurons, 20 collocation points, and a time-step of length 0.1. The results with this setup are reported in Table \ref{tab:BBXTFC_lorenz_20_20_0.1}, where we see a worse performance for the learned dynamics, but a better one for the RHS, whose MAE can achieve the order of 1e-09. We choose the RHS learned with this setup for our symbolic regression distillations, as it will lead to more accurate results, as reported in the next subsection. Notice that, in order to perform the regression, the number of collocation points of BBX-TFC through the whole domain needs to be the same as the number of data points. To do this, the user can manually select the sub-domain length and the number of collocation points per sub-domain accordingly or automatically adjust the dataset for a fixed framework's setup with a classical polynomial interpolation. The latter allows greater freedom in parameter selection without interfering with the performance.\\ 
\begin{table}[h!]
\centering
\begin{tabular}{lccc|ccc}
\hline
\multicolumn{1}{c}{} &  \multicolumn{6}{c}{\textbf{MAE}}  \\ \hline
\multicolumn{1}{c}{\textbf{data points}}  & \multicolumn{1}{c}{\boldsymbol{$x$}}  & \multicolumn{1}{c}{\boldsymbol{$y$}}  & \multicolumn{1}{c|}{\boldsymbol{$z$}} & \multicolumn{1}{c}{\boldsymbol{$\dot x$}} &  \multicolumn{1}{c}{\boldsymbol{$\dot y$}} & \multicolumn{1}{c}{\boldsymbol{$\dot z$}}    \\ \hline \hline
200 & 1.04e-12 & 3.15e-12 & 2.31e-12 & 1.69e-01 & 5.31e-01 & 4.95e-01 \\
500 & 1.20e-13 & 3.47e-13 & 3.59e-13 & 1.88e-02 & 6.01e-02 & 5.28e-02 \\
1000 & 1.57e-15 & 3.22e-15 & 4.29e-15 & 1.91e-04 & 3.97e-04 & 3.35e-04 \\
2000  & 2.82e-15 & 1.01e-14 & 7.18e-15 & 3.44e-05 & 1.23e-04 & 1.02e-04 \\
5000  & 7.24e-16 & 1.82e-15 & 1.47e-15 & 1.17e-04 & 3.38e-04 & 3.18e-04 \\
10000  & 2.32e-15 & 8.90e-15 & 5.68e-15 & 1.17e-04 & 4.53e-04 & 3.97e-04 \\
20000 & 8.26e-16 & 2.43e-15 & 1.98e-15 & 4.19e-05 & 1.21e-04 & 1.16e-04 \\
50000 & 1.24e-15 & 2.10e-15 & 2.19e-15 & 1.20e-04 & 3.10e-04 & 3.01e-04 \\
\hline
\hline
\end{tabular}
\caption{Lorenz system: Black-Box X-TFC performance for learned dynamics and learned RHS in terms of MAE, by varying the number of data points, by using 10 neurons, 10 collocation points per sub-domain, and a subdomain length of 0.1.}\label{tab:BBXTFC_lorenz_10_10_0.1}
\end{table}
\begin{table}[h!]
\centering
\begin{tabular}{lccc|ccc}
\hline
\multicolumn{1}{c}{} &  \multicolumn{6}{c}{\textbf{MAE}}  \\ \hline
\multicolumn{1}{c}{\textbf{data points}}  & \multicolumn{1}{c}{\boldsymbol{$x$}}  & \multicolumn{1}{c}{\boldsymbol{$y$}}  & \multicolumn{1}{c|}{\boldsymbol{$z$}} & \multicolumn{1}{c}{\boldsymbol{$\dot x$}} &  \multicolumn{1}{c}{\boldsymbol{$\dot y$}} & \multicolumn{1}{c}{\boldsymbol{$\dot z$}}    \\ \hline \hline
200 & 3.68e-20 & 6.62e-20 & 0.00 & 1.50e-01 & 4.59e-01 & 4.42e-01 \\ 
500 & 2.73e-20 & 3.12e-20 & 0.00 & 5.93e-03 & 1.77e-02 & 1.59e-02 \\ 
1000 & 2.12e-19 & 1.42e-19 & 0.00 & 6.15e-04 & 1.67e-03 & 1.57e-03 \\ 
2000  & 2.71e-20 & 2.94e-20 & 0.00 & 7.21e-05 & 1.95e-04 & 1.85e-04 \\
5000  & 9.52e-20 & 9.84e-20 & 0.00 & 1.71e-05 & 2.76e-05 & 3.35e-05 \\ 
10000  & 3.04e-20 & 5.19e-20 & 0.00 & 3.02e-06 & 4.54e-06 & 5.75e-06 \\ 
20000 & 2.68e-20 & 4.58e-20 & 0.00 & 1.49e-05 & 1.93e-05 & 2.61e-05 \\
50000 & 1.60e-19 & 1.94e-19 & 0.00 & 3.86e-05 & 4.98e-05 & 6.76e-05 \\ 
\hline
\hline
\end{tabular}
\caption{Lorenz system: Black-Box X-TFC performance for learned dynamics and learned right-hand-sides in terms of MAE, by varying the number of data points, using 10 neurons, 10 collocation points per sub-domain, and subdomain length of 0.001.}\label{tab:BBXTFC_lorenz_10_10_0.001}
\end{table}
\begin{table}[h!]
\centering
\begin{tabular}{lccc|ccc}
\hline
\multicolumn{1}{c}{} &  \multicolumn{6}{c}{\textbf{MAE}}  \\ \hline
\multicolumn{1}{c}{\textbf{data points}}  & \multicolumn{1}{c}{\boldsymbol{$x$}}  & \multicolumn{1}{c}{\boldsymbol{$y$}}  & \multicolumn{1}{c|}{\boldsymbol{$z$}} & \multicolumn{1}{c}{\boldsymbol{$\dot x$}} &  \multicolumn{1}{c}{\boldsymbol{$\dot y$}} & \multicolumn{1}{c}{\boldsymbol{$\dot z$}}    \\ \hline \hline
200 &  1.33e-06 & 3.88e-06 & 3.20e-06 & 1.73e-01 & 5.33e-01 & 5.01e-01 \\
500 & 9.13e-07 & 2.80e-06 & 2.87e-06 & 1.11e-02 & 3.00e-02 & 2.95e-02 \\ 
1000 & 4.30e-07 & 1.22e-06 & 1.17e-06 & 1.77e-03 & 7.11e-03 & 5.04e-03 \\ 
2000  & 6.68e-11 & 2.04e-10 & 9.24e-11 & 1.14e-05 & 2.24e-05 & 1.75e-05 \\ 
5000  & 5.68e-10 & 1.44e-09 & 1.49e-09 & 7.26e-06 & 1.82e-05 & 1.63e-05 \\ 
10000  & 8.31e-12 & 2.99e-11 & 2.27e-11 & 9.60e-08 & 3.44e-07 & 2.85e-07 \\ 
20000 & 2.21e-12 & 6.24e-12 & 5.99e-12 & 1.82e-08 & 5.27e-08 & 4.20e-08 \\ 
50000 & 4.30e-14 & 1.44e-13 & 1.49e-13 & 2.44e-09 & 8.66e-09 & 8.99e-09 \\
\hline
\hline
\end{tabular}
\caption{Lorenz system: Black-Box X-TFC performance for learned dynamics and learned right-hand-sides in terms of MAE, by varying the number of data points, using 20 neurons, 20 collocation points per sub-domain, and subdomain length of 0.1.}\label{tab:BBXTFC_lorenz_20_20_0.1}
\end{table}

To complete our study on the BBX-TFC performance, we report the errors of the RHS directly learned with Neural Networks approximating the missing terms $f(x,y,z)$, $h(x,y,z)$, and $k(x,y,z)$ of the system of ODEs
\begin{align*}
    \dot x &= f (x,y,z)\\
    \dot y &= h(x,y,z) \\
    \dot z &= k (x,y,z)
\end{align*}
as shown in the Methodology section \ref{sec:methodology} above. The results in Table \ref{tab:BBXTFC_lorenz_fun_20_20_0.1} are obtained with 20 neurons, 20 collocation points, and a time-step of length 0.1, show that -- with this parameters setup -- we can find negligible improvements in terms of MAE for the scenarios with smaller datasets (from 200 to 2000 data points). However, the previous (indirect) framework still outperforms the present one for the other datasets, in both accuracy and computational times (for example, for the case with 10000 data points, computational times of 1.5 seconds vs. 6.5 seconds).
\begin{table}[h!]
\centering
\begin{tabular}{lccc|ccc|ccc}
\hline
\multicolumn{1}{c}{} &  \multicolumn{9}{c}{\textbf{MAE}}  \\ \hline
\multicolumn{1}{c}{\textbf{data points}}  & \multicolumn{1}{c}{\boldsymbol{$x$}}  & \multicolumn{1}{c}{\boldsymbol{$y$}}  & \multicolumn{1}{c|}{\boldsymbol{$z$}} & \multicolumn{1}{c}{\boldsymbol{$\dot x$}} &  \multicolumn{1}{c}{\boldsymbol{$\dot y$}} & \multicolumn{1}{c|}{\boldsymbol{$\dot z$}} & \multicolumn{1}{c}{\boldsymbol{$f$}} &  \multicolumn{1}{c}{\boldsymbol{$h$}} & \multicolumn{1}{c}{\boldsymbol{$k$}}    \\ \hline \hline
200 & 6.31e-06 & 1.85e-05 & 1.52e-05 & 1.53e-01 & 4.69e-01 & 4.50e-01 & 1.53e-01 & 4.69e-01 & 4.50e-01 \\
500 & 6.41e-06 & 1.64e-05 & 1.68e-05 & 9.71e-03 & 2.72e-02 & 2.52e-02 & 9.71e-03 & 2.72e-02 & 2.52e-02 \\ 
1000 & 6.27e-07 & 1.94e-06 & 1.74e-06 & 4.34e-04 & 1.29e-03 & 9.39e-04 & 4.34e-04 & 1.29e-03 & 9.39e-04 \\
2000  & 3.56e-10 & 1.07e-09 & 5.21e-10 & 9.20e-06 & 1.62e-05 & 1.46e-05 & 9.20e-06 & 1.62e-05 & 1.46e-05 \\ 
5000  & 1.12e-09 & 2.82e-09 & 2.94e-09 & 3.17e-07 & 1.12e-06 & 7.99e-07 & 3.17e-07 & 1.12e-06 & 7.99e-07 \\
10000  & 5.00e-11 & 1.42e-10 & 1.68e-10 & 8.88e-08 & 2.39e-07 & 2.89e-07 & 8.88e-08 & 2.39e-07 & 2.89e-07 \\ 
20000 & 1.75e-10 & 3.94e-10 & 4.75e-10 & 1.49e-07 & 4.31e-07 & 4.11e-07 & 1.49e-07 & 4.31e-07 & 4.11e-07 \\ 
50000 & 7.47e-12 & 1.89e-11 & 2.29e-11 & 3.27e-08 & 8.10e-08 & 1.00e-07 & 3.27e-08 & 8.10e-08 & 1.00e-07 \\ 
\hline
\hline
\end{tabular}
\caption{Lorenz system: Black-Box X-TFC performance for learned dynamics and for indirect and direct learned RHS in terms of MAE, by varying the number of data points, by using 20 neurons, 20 collocation points per sub-domain, and subdomain length of 0.1.}\label{tab:BBXTFC_lorenz_fun_20_20_0.1}
\end{table}
We generated the datasets at 20 Hz up to 5000 Hz to study the regression performance at different degrees of data sparsity. In Ref. \cite{sun2023pisl}, the authors discover the Lorenz system with a sparse dataset generated in [0,20] seconds at 20 Hz. Goyal and Benner \cite{goyal2022discovery} integrated the classical 4th-order Runge-Kutta with SINDy to discover the Lorenz system with sparse data generated at 100 Hz. Raissi et al. \cite{raissi2018multistep} extracted the nonlinear dynamics of Lorenz attractor from time-series at 100 Hz by merging classical multi-step time-stepping schemes with deep neural networks as nonlinear function approximators.


\subsubsection{PySR performance}

%

\begin{figure}[h!]
    \centering
    \begin{subfigure}{\textwidth}
        \centering
        \includegraphics[width=\linewidth, trim=120 0 120 0, clip]{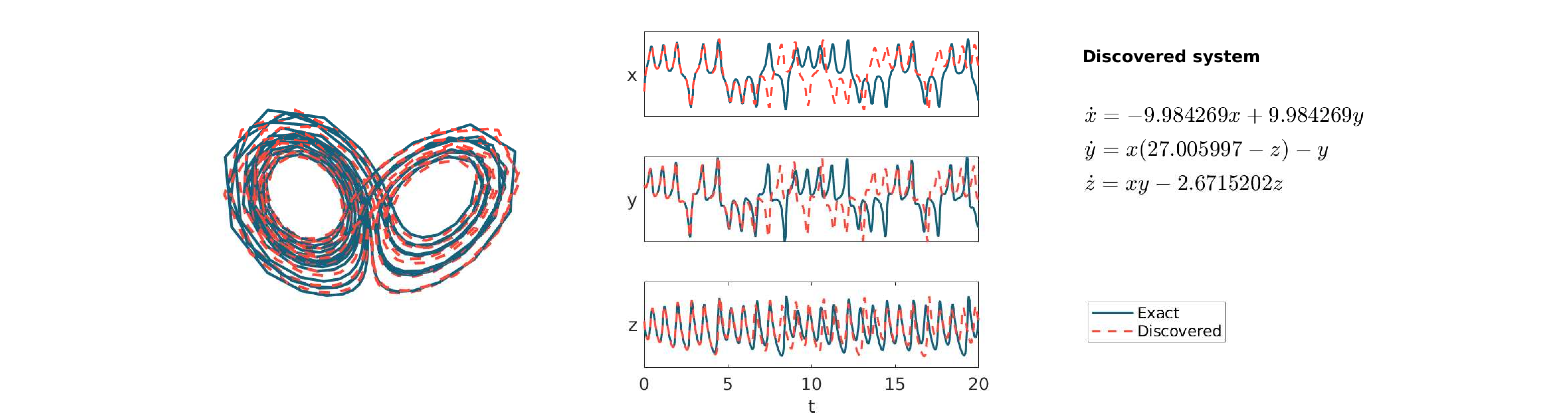}
        \caption{PySR discovered dynamics vs. exact dynamics using Black-Box X-TFC for 200 points.}
        \label{fig:pysr_lorenz_200}
    \end{subfigure}

    \begin{subfigure}{\textwidth}
        \centering
        \includegraphics[width=\linewidth, trim=120 0 120 0, clip]{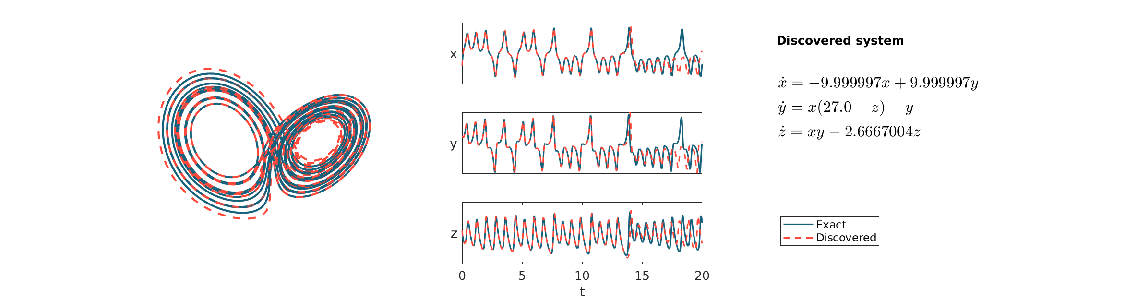}
        \caption{PySR discovered dynamics vs. exact dynamics using Black-Box X-TFC for 2000 points.}
        \label{fig:pysr_lorenz_2000}
    \end{subfigure}

    \begin{subfigure}{\textwidth}
        \centering
        \includegraphics[width=\linewidth, trim=120 0 120 0, clip]{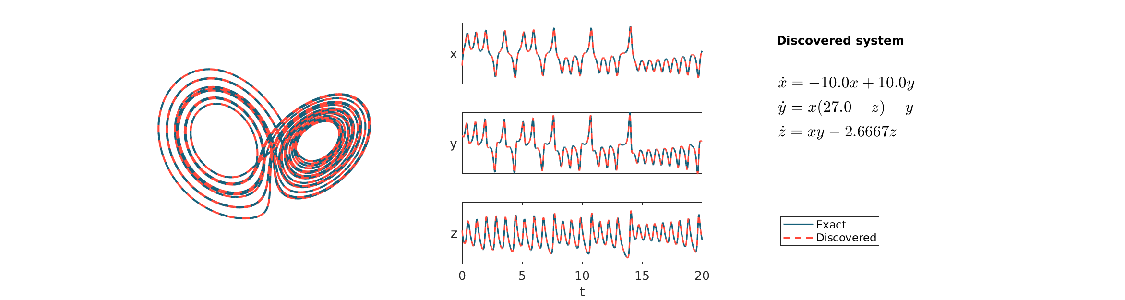}
        \caption{PySR discovered dynamics vs. exact dynamics using Black-Box X-TFC for 5000 points.}
        \label{fig:pysr_lorenz_5000}
    \end{subfigure}
    \caption{Lorenz system: Comparison of PySR discovered dynamics from Black-Box X-TFC for two different sets of data points. The discovered RHS terms of the differential equations are shown on the right.}
    \label{fig:pysr_comparison_lorenz}
\end{figure}

The Black-Box X-TFC outputs can now be used as inputs in the Symbolic Regression framework, PySR. With PySR, we aim to find the mathematical expressions made by linear and nonlinear combinations of $x$, $y$, and $z$ that best fit the functions $\dot x$, $\dot y$, and $\dot z$. Unlike the SINDy algorithm, PySR does not require a library of the possible terms that can appear in the discovered equation. Instead, we only need to declare a library of certain binary operators, e.g., $+, -, *, /$, unary operators, e.g., \textit{cube, exp, log, abs}, and tuning hyperparameters such as population number and number of iterations. For the test cases in this study, we only need to declare the binary operators $+,-$, and $*$, and set the population number and number of iterations equal to 30. The accuracy of the expressions and the precision of the parameters discovered by PySR depends on the quality of the input functions provided to it, which depends on the quality of the input data used for the Black-Box regression. This can be seen in Figure \ref{fig:pysr_comparison_lorenz}, where the PySR performance is reported by using the dynamics learned with 200 data points (\ref{fig:pysr_lorenz_200}), 2000 data points (\ref{fig:pysr_lorenz_2000}), and 5000 data points (\ref{fig:pysr_lorenz_5000}), and by using the learned RHS of Table \ref{tab:BBXTFC_lorenz_20_20_0.1}. The mathematical expressions distilled by PySR are reported on the right side of the plots. In the plots, the blue line represents the exact solution of the Lorenz system computed by the Runge-Kutta method with a certain selected time step. For example, in the first case, the time-step size is 0.05, which generates a dataset of 200 points. The red dashed line is obtained using the Runge-Kutta method to solve the discovered equations, selecting the same time-step used for the data generation. For all cases, the exact equations are discovered, with the difference in the precision of the parameters' values, which will obviously lead to a separation between the predicted and the exact trajectory at a certain point in time, given the chaotic nature of the dynamical system and its sensitivity to small perturbations. In particular, for the case with 200 data points, we are able to predict the Lorenz trajectory up to about 8 seconds, and further up to about 17 seconds with 2000 data points. When we have enough data points, from 5000 points onwards, the PySR algorithm is able to precisely identify the exact value of the parameters of the differential equations. 

While comparing time-series predictions between the original and the identified system is a simple,  frequently used and appealing approach for visually establishing success of the identification process, what we aim at is the accurate identification of the right-hand-side of the unavailable equations -- a bound for the norm of the difference between two functions in state space.
Because of the sensitivity of chaotic system trajectories to initial conditions, depending on the system's positive Lyapunov exponents one may appear to be more or less successful in exactly matching chaotic time series over finite times; we present overlays of the corresponding attractors  as an alternative informative visual comparison. Still, one should note that the lack of robustness of the chaotic attractor details to infinitesimal parameter changes, so that small changes in the identified parameters might, some times, give rise to strikingly different long term attractors (e.g. at the boundaries of intermittency) even though the identified right-hand-side is very close to the true one in any reasonable norm.

We present a comparison between PySR and another symbolic regressor named \textit{gplearn} (Genetic Programming for Symbolic Regression) \cite{stephens2015gplearn}. The gplearn package, as for PySR, is a genetic programming based algorithm implemented in Python, which initializes a random population of candidate solutions, evaluated based on how well they fit the data. The candidates with the best fitness are selected and go through crossover and mutations, creating new candidates for the next generations. This process terminates when certain conditions are satisfied, or when a maximum number of iterations is reached. In Table \ref{tab:comparison_sr}, the right-hand sides of the Lorenz system discovered by PySR (left) and gplearn (right) are reported, for datasets of 200, 2000, and 5000 observed points. The gplearn results are obtained using an initial population of 5000 candidates and 200 iterations. One can see that gplearn, for each dataset, is able to discover the correct RHS of the first differential equation with quite good precision in the parameter discovery. For the 200 data points case, gplearn also fails to learn the second and third RHS but improves its performance for the 5000 data points case, in which two equations (first and second) are correctly discovered. These results demonstrate why PySR, for our test case, proved to be the best choice of symbolic regressor.

\begin{table}[h!]
\centering
\begin{tabular}{ll|l}
 \hline
\multicolumn{1}{c}{} &  \multicolumn{2}{c}{\textbf{Discovered equations with symbolic regressors}}  \\ \hline
\multicolumn{1}{c}{\textbf{data points}}  & \multicolumn{1}{c|}{\textbf{PySR}} & \multicolumn{1}{c}{\textbf{gplearn}} \\ \hline \hline
    & $\dot x = 9.984269(y-x)$  & $\dot x = 9.975373  (y-x)$   \\
    200 & $\dot y = x (27.005997 - z) - y$  &  $\dot y = 2.63852 x - [(4.84091 x^3) + (1.14025 y) + (2.84091 x^2 y)]/z $  \\
    &  $\dot z = xy - 2.6715202 z$ & $\dot z = xy - 2.647446 z - 0.394 $ \\      
    \hline
    &  $\dot x = 9.999997(y-x)$  &  $\dot x = 9.984899 (y - x) $  \\
2000 &  $\dot y = x(27 - z) - y$ & $\dot y = 
24.6248 x - 0.949932 x z $ \\
    &  $\dot z = xy - 2.6667004z$  &   $\dot z = x y - 2.65289 z - 0.307 $ \\   
   \hline
    &  $\dot x = 10(y-x)$  &  $\dot x =  9.984899 (y - x) $  \\   
5000 & $\dot y = x(27 - z) - y $  &   $\dot y = 26.918 x -0.994719 x z   - y $ \\
    & $\dot z = xy - 2.6667 z$   &  $\dot z = x y - 2.65289 z - 0.307 $  \\   
\hline
\hline
\end{tabular}
\caption{Lorenz system: Comparison between PySR and gplearn performance for equations discovery, using 200, 2000, and 5000 observed data points.}\label{tab:comparison_sr}
\end{table}


\subsubsection{Comparison with Gaussian Process Regression plus PySR}

\begin{figure}[h!]
    \centering
    \begin{subfigure}{\textwidth}
        \centering
        
    \end{subfigure}
    \begin{subfigure}{\textwidth}
        \centering
        \includegraphics[width=\linewidth, trim=120 0 0 0, clip]{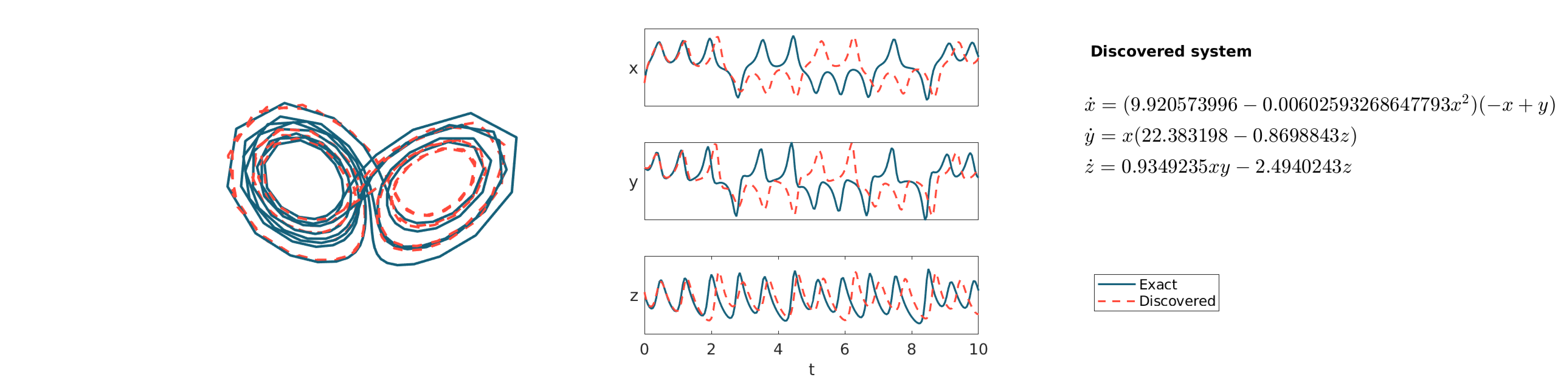}
        \caption{PySR discovered dynamics vs. exact dynamics using Gaussian Process Regression with 200 data points.}
        \label{fig:pysr_lorenz_gpr_200}
        \includegraphics[width=\linewidth, trim=120 0 0 0, clip]{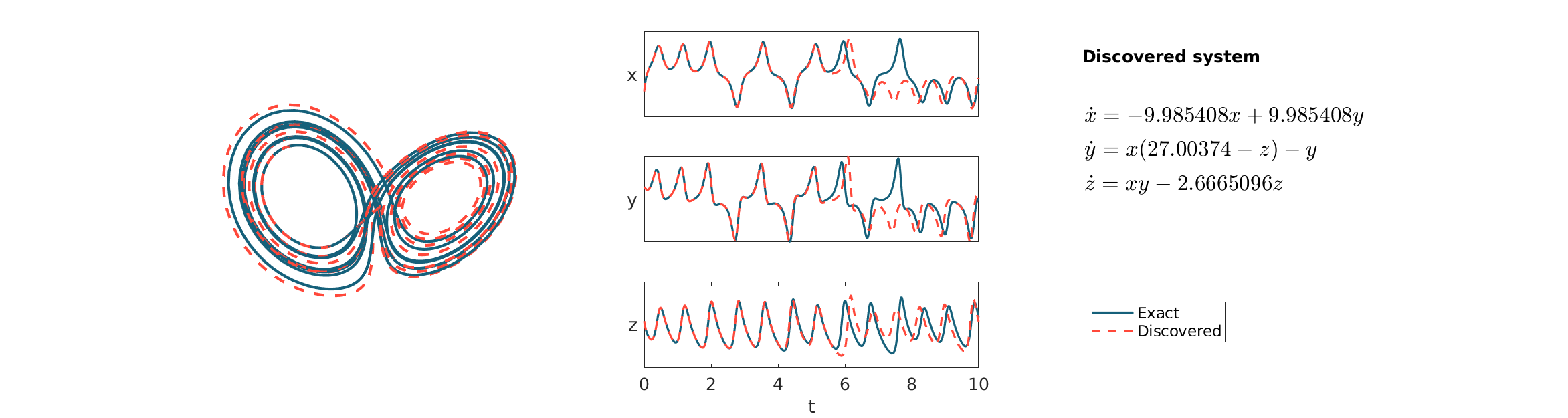}
        \caption{PySR discovered dynamics vs. exact dynamics using Gaussian Process Regression with 1000 data points.}
        \label{fig:pysr_lorenz_gpr_1000}
        \includegraphics[width=\linewidth, trim=120 0 0 0, clip]{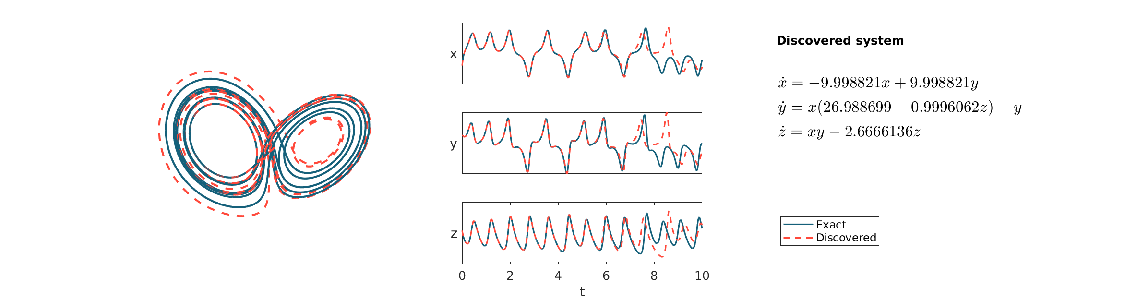}
        \caption{PySR discovered dynamics vs. exact dynamics using Gaussian Process Regression with 2000 data points.}
        \label{fig:pysr_lorenz_gpr_2000}
    \end{subfigure}
    \caption{Lorenz System: Comparison of PySR discovered dynamics from Gaussian Process Regression for three different sets of data points. The discovered RHS terms of the differential equations are shown on the right.}
    \label{fig:pysr_lorenz_gpr}
\end{figure}

Next, we compare the Black-Box X-TFC with Gaussian Process Regression (GPR) \cite{williams1995gaussian,rasmussen2003gaussian,seeger2004gaussian}. The GPR algorithm is implemented with the ``GaussianProcessRegressor" package from the \textit{sklearn.gaussian\_process} library in Python \cite{williams2006gaussian}. Once the regression of the dynamics data is performed, the right-hand-side functions are computed by numerically differentiating the learned dynamics. In Table \ref{tab:comparison_tfc_gpr}, we present the performance of BBX-TFC and GPR for regression tasks, for different numbers of data points, from 200 to 2000. The comparison is given in terms of MAE and computational times. For all test cases, BBX-TFC outperforms GPR in terms of both error and computational efficiency. Due to its accuracy in performing regression, BBX-TFC is a great candidate tool to feed PySR and obtain precise differential equations. In Figure \ref{fig:pysr_lorenz_gpr}, the PySR discovered equations obtained using the regression outputs from GPR for 200, 1000, and 2000 points, are plotted in \ref{fig:pysr_lorenz_gpr_200}, \ref{fig:pysr_lorenz_gpr_1000} and \ref{fig:pysr_lorenz_gpr_2000}, respectively. For the case with 200 points, the discovered equations are not exact, except for the third one, leading to a prediction of the trajectory for 2 seconds. For the case with 1000 points, PySR is able to learn the exact equations, but lacking in parameters precision, leading to a deviation of the predicted trajectory from the exact one after 6 seconds. Finally, for the case with 2000 data points, the precision in finding the governing parameters improves, leading to a good prediction of the Lorenz trajectory up to 7 seconds.
\begin{table}[h!]
\centering
\begin{tabular}{lccc|ccc|c}
 \hline
\multicolumn{1}{c}{} &  \multicolumn{6}{c}{\textbf{GPR}}  \\ \hline
\multicolumn{1}{c}{\textbf{data points}}  & \multicolumn{1}{c}{\boldsymbol{$x$}}  & \multicolumn{1}{c}{\boldsymbol{$y$}}  & \multicolumn{1}{c|}{\boldsymbol{$z$}} & \multicolumn{1}{c}{\boldsymbol{$\dot x$}} &  \multicolumn{1}{c}{\boldsymbol{$\dot y$}} & \multicolumn{1}{c}{\boldsymbol{$\dot z$}}  & comp. time  \\ \hline \hline
200 & 4.88e-08 &  3.78e-09  & 6.32e-09 & 2.25e-00 & 4.47e-00 & 4.40e-00 & 6.6 sec. \\
500 &  2.57e-06 & 2.59e-06 & 2.79e-06 & 3.88e-01 & 7.99e-01 & 8.04e-01 & 40 sec. \\
1000 & 1.83e-06  &  1.64e-06 &  1.73e-06 &  9.94e-02 & 2.00e-01  &  1.96e-01   &  2 min. \\
2000 & 1.30e-06  &  1.15e-06 &  1.21e-06 &  2.49e-02 & 4.99e-02  &  4.90e-02  &  8 min. \\
\hline \hline
\multicolumn{1}{c}{} &  \multicolumn{6}{c}{\textbf{BBX-TFC}}  \\ \hline
\multicolumn{1}{c}{\textbf{data points}}  & \multicolumn{1}{c}{\boldsymbol{$x$}}  & \multicolumn{1}{c}{\boldsymbol{$y$}}  & \multicolumn{1}{c|}{\boldsymbol{$z$}} & \multicolumn{1}{c}{\boldsymbol{$\dot x$}} &  \multicolumn{1}{c}{\boldsymbol{$\dot y$}} & \multicolumn{1}{c}{\boldsymbol{$\dot z$}}  & comp. time  \\ \hline \hline
200 &  1.33e-06 & 3.88e-06 & 3.20e-06 & 1.73e-01 & 5.33e-01 & 5.01e-01 & 2.3 sec. \\
500 & 9.13e-07 & 2.80e-06 & 2.87e-06 & 1.11e-02 & 3.00e-02 & 2.95e-02  & 2.3 sec. \\ 
1000 & 4.30e-07 & 1.22e-06 & 1.17e-06 & 1.77e-03 & 7.11e-03 & 5.04e-03  & 2.3 sec. \\ 
2000  & 6.68e-11 & 2.04e-10 & 9.24e-11 & 1.14e-05 & 2.24e-05 & 1.75e-05  & 1.1 sec. \\ 
\hline
\hline
\end{tabular}
\caption{Lorenz system: Comparison between Black-Box X-TFC and Gaussian Process Regression performance for learned dynamics and learned RHS in terms of MAE and computational time, for number of data points in a range [200,2000].}\label{tab:comparison_tfc_gpr}
\end{table}

\clearpage

\subsubsection{Noisy data}

\begin{figure}[h!]
    \centering
    \begin{subfigure}{\textwidth}
        \centering
        \includegraphics[width=0.8\linewidth]{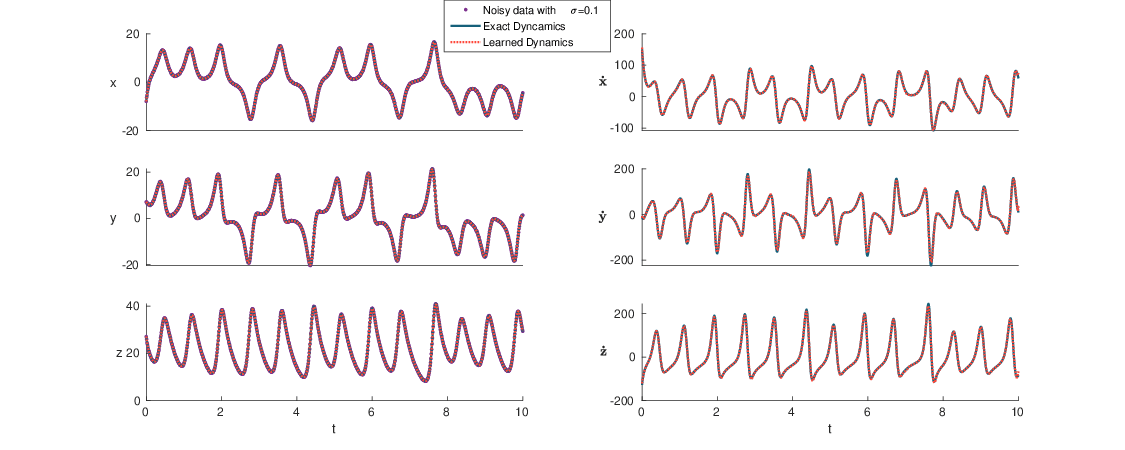}
        \caption{Noise standard deviation $\sigma=0.1$.}
        \label{fig:bbxtfc_lorenz_noise_01}
    \end{subfigure}
    
    \begin{subfigure}{\textwidth}
        \centering
        \includegraphics[width=0.8\linewidth]{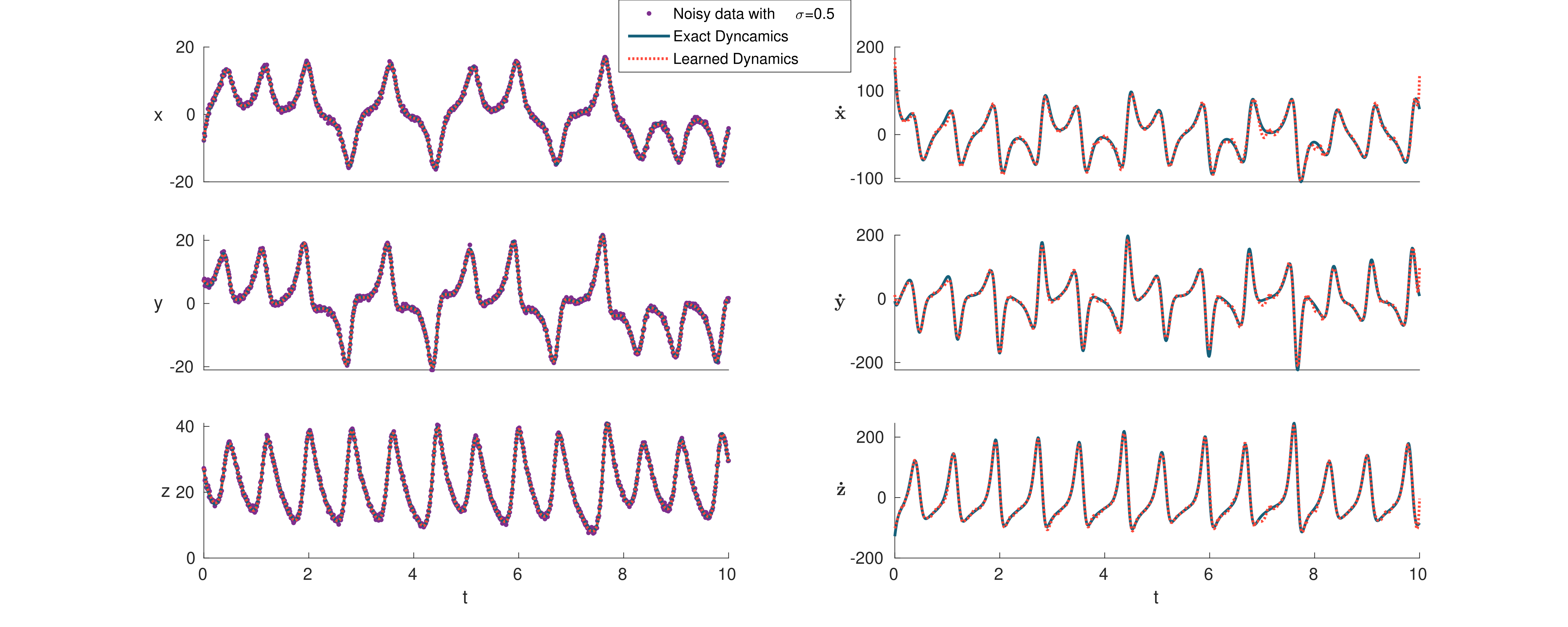}
        \caption{Noise standard deviation $\sigma=0.5$.}
        \label{fig:bbxtfc_lorenz_noise_05}
    \end{subfigure}

    \begin{subfigure}{\textwidth}
        \centering
        \includegraphics[width=0.8\linewidth]{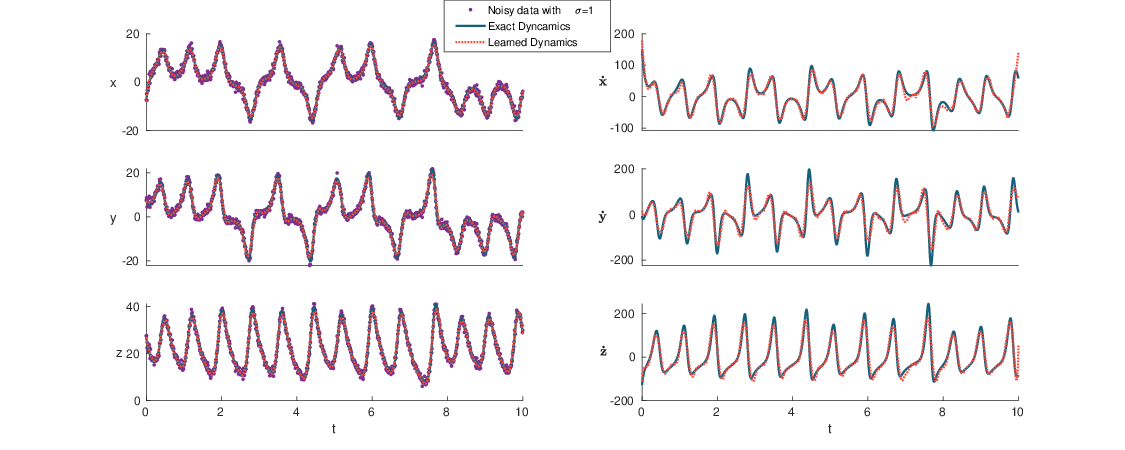}
        \caption{Noise standard deviation $\sigma=1$.}
        \label{fig:bbxtfc_lorenz_noise_1}
    \end{subfigure}
    
    \begin{subfigure}{\textwidth}
        \centering
        \includegraphics[width=0.8\linewidth]{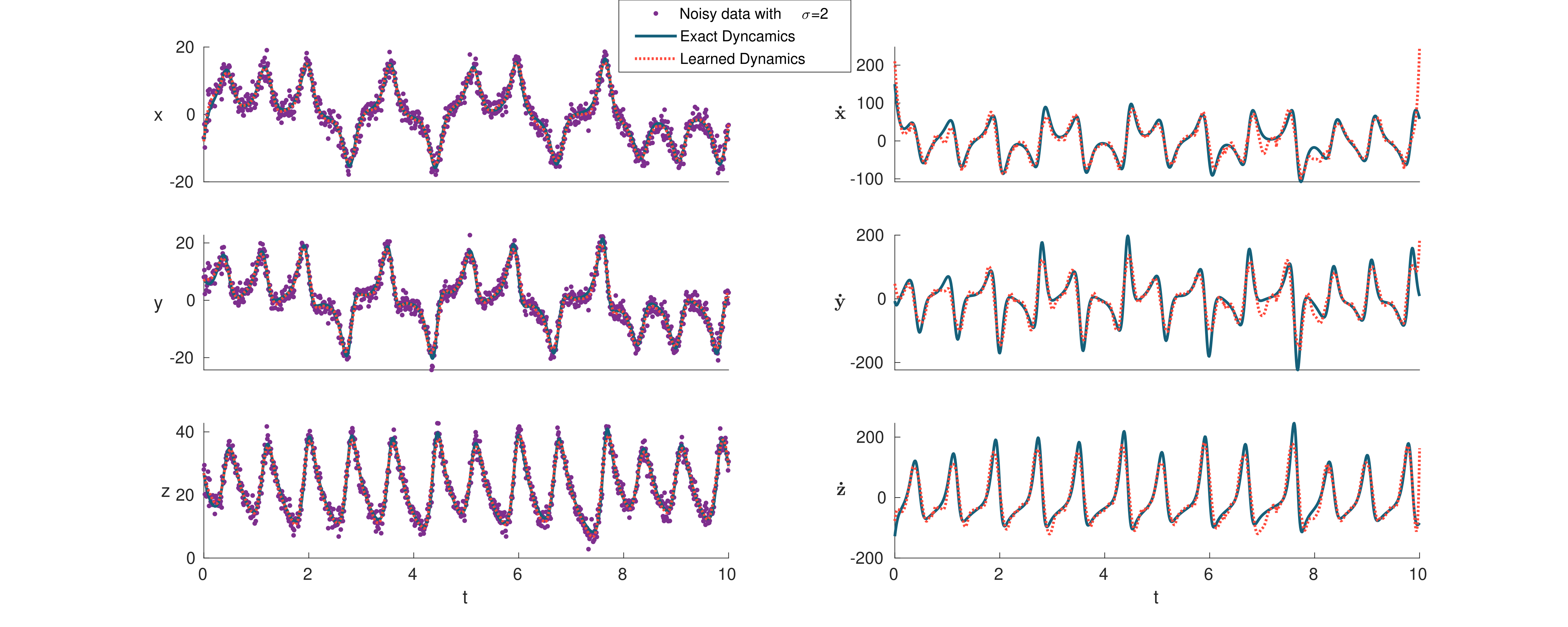}
        \caption{Noise standard deviation $\sigma=2$.}
        \label{fig:bbxtfc_lorenz_noise_2}
    \end{subfigure}

    \caption{Lorenz system: Black-Box X-TFC learned dynamics (left) and right-hand-side (right) vs. exact solutions, obtained with 1000 data points with different levels of noise.}
    \label{fig:bbxtfc_lorenz_noise}
\end{figure}

\begin{figure}[h!]
    \centering
    \begin{subfigure}{\textwidth}
        \centering
        \includegraphics[width=0.8\linewidth, trim=100 0 100 0, clip]{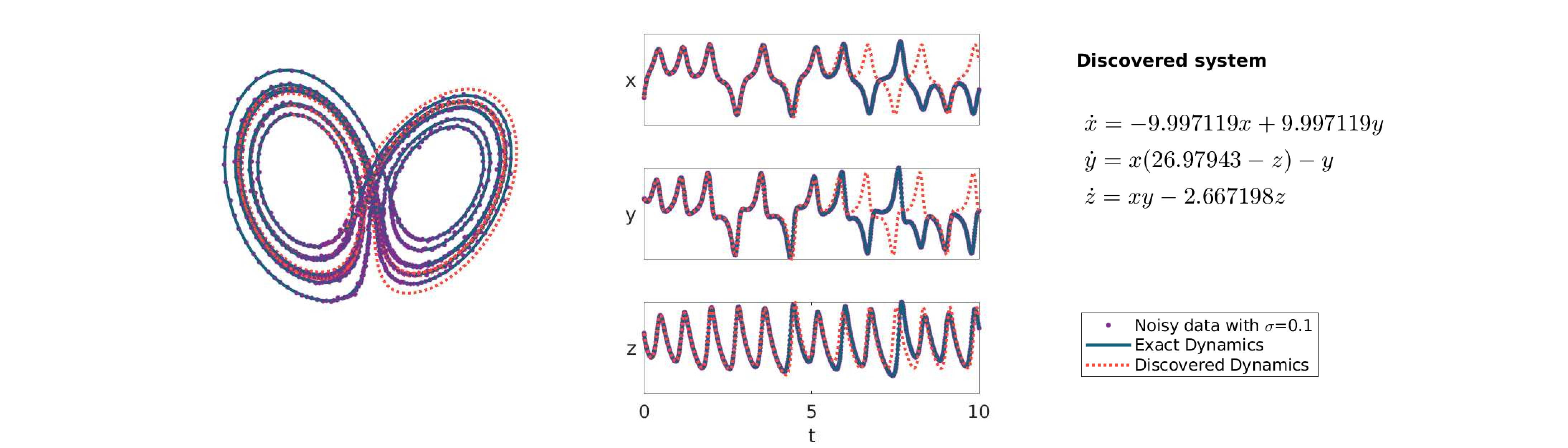}
        \caption{Noise standard deviation $\sigma=0.1$.}
        \label{fig:SR_lorenz_noise_01}
    \end{subfigure}
    
    \begin{subfigure}{\textwidth}
        \centering
        \includegraphics[width=0.8\linewidth, trim=100 0 100 0, clip]{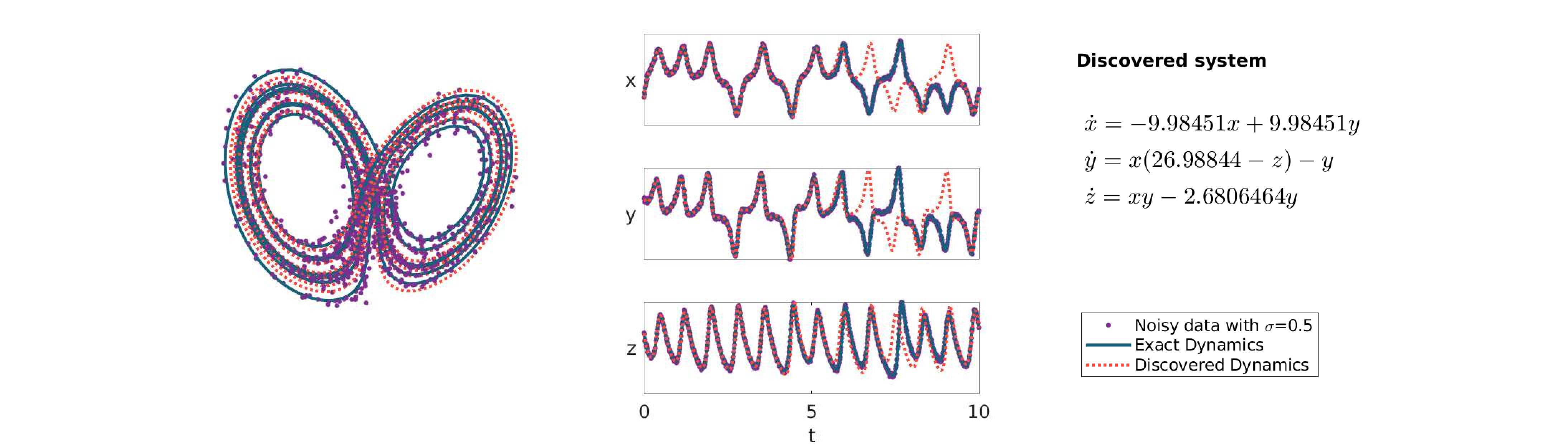}
        \caption{Noise standard deviation $\sigma=0.5$.}
        \label{fig:SR_lorenz_noise_05}
    \end{subfigure}

    \begin{subfigure}{\textwidth}
        \centering
        \includegraphics[width=0.8\linewidth, trim=100 0 100 0, clip]{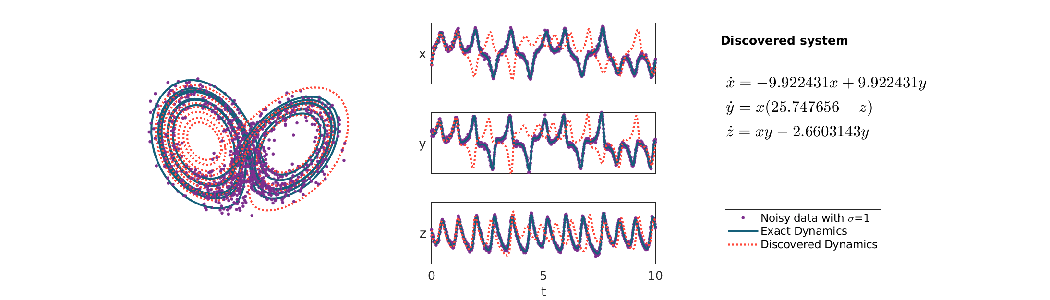}
        \caption{Noise standard deviation $\sigma=1$.}
        \label{fig:SR_lorenz_noise_1}
    \end{subfigure}
    
    \begin{subfigure}{\textwidth}
        \centering
        \includegraphics[width=0.8\linewidth, trim=100 0 100 0, clip]{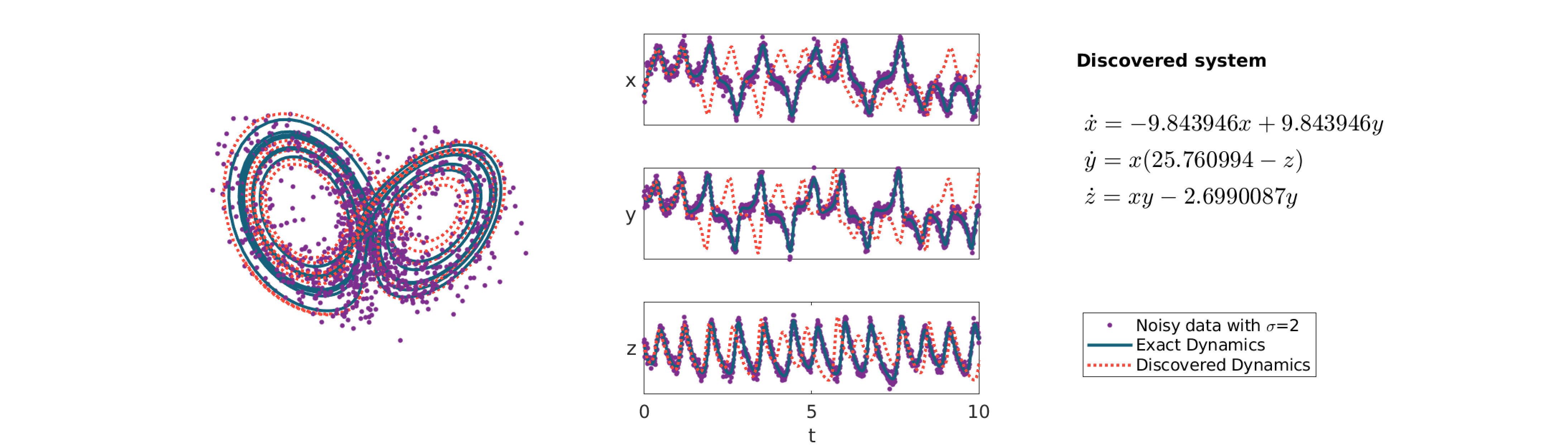}
        \caption{Noise standard deviation $\sigma=2$.}
        \label{fig:SR_lorenz_noise_2}
    \end{subfigure}

    \caption{Lorenz system with noisy data: comparison of PySR discovered dynamics from Black-Box X-TFC for 1000 noisy data points, with noise standard deviation $\sigma = [0.1,0.5,1.0,2.0]$. The discovered RHS terms of the differential equations are shown on the right.}
    \label{fig:SR_lorenz_noise}
\end{figure}

The presence of noise in the measurement data still represents a fundamental challenge for model discovery methods, which rely on the accuracy of the derivatives computation \cite{brunton2016discovering,galioto2020bayesian}. To this end, we contaminate our data with a Gaussian distribution noise with mean $\mu=0$ and different values of standard deviation $\sigma = [0.1 , 0.5 , 1.0 , 2.0]$, giving four different levels of noise, to test the robustness of our framework for a more realistic scenario. To achieve the performance shown in Figure \ref{fig:bbxtfc_lorenz_noise}, BBX-TFC needs to be tuned with a larger time-step size in order to avoid overfitting. The learned RHS present outlier values, which can be smoothed by using a Savitzky-Golay filtering, to improve the performance of PySR. Herein, we use a Savitzky-Golay filter with polynomial order 5, and frame length of 31 and 55, for noise $\sigma = [0.1 , 0.5]$ and $\sigma = [1.0 , 2.0]$, respectively. 
Figure \ref{fig:SR_lorenz_noise} shows the discovered equations for the Lorenz system with noisy data at different levels of $\sigma$. 
When noise is relatively low ($\sigma = [0.1,0.5]$), the SR algorithm perfectly discovers the differential equations with good precision in the parameters. For both cases, the predicted trajectory is accurate up to 6 seconds. When the level of noise becomes higher ($\sigma = [1.0,2.0]$), PySR is able to discover the right form of the first and third equation and part of the second one, missing the term containing the presence of $y$. For both cases, we have a good trajectory prediction of up to 3 seconds, before it deviates from the exact trajectory, but still being able to reproduce the behavior of the Lorenz dynamical system and its butterfly shape.

\clearpage


\subsubsection{Gray-Box X-TFC and PySR performance}

\begin{figure}[h!]
    \centering
    \includegraphics[width=0.7\linewidth]{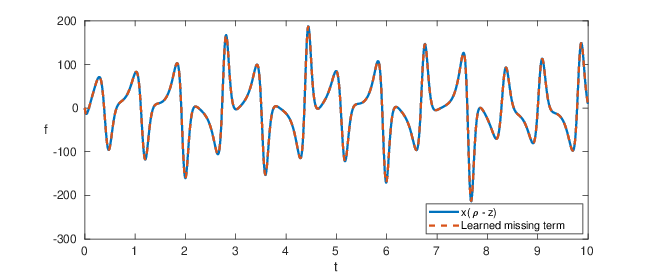}
    \caption{Missing term identification in the Lorenz system: Gray-Box X-TFC learned unknown term vs. exact solution.}
    \label{fig:gray_xtfc_lorenz}
\end{figure}

To conclude with the Lorenz system test case, we briefly show the performance of Gray-Box X-TFC and PySR to tackle gray-box problems when partial knowledge of the physical system is available. The retrieved unknown term $f(x,y,z)$, described in the system of equations \eqref{eq:gray_lorenz}, is plotted in Figure \ref{fig:gray_xtfc_lorenz}, showing a good match compared with the exact value of the differential equation term. 

The Runge-Kutta method generated the dataset used in this example with time-step 0.01 in a time domain $[0,10]$, resulting in 1000 data points. The GBX-TFC performed with 20 collocation points per sub-domain, 20 neurons, and a sub-domain length of 0.1. The MAE between the exact and learned missing terms is $7.09e-03$, with a computational time of 3.5 seconds. By feeding the PySR algorithm with the learned unknown function, we can obtain its mathematical expression, which will complete the physical knowledge of the Lorenz system. In this case, the SR output is
\begin{equation}
    f = x (26.999224 - z)
\end{equation}
which agrees with the exact term of the differential equation. \\
Table \ref{tab:GBXTFC_lorenz_fun_10_10_0.1} shows the performance of GBX-TFC in terms of MAE for the learned missing term in the Lorenz system. These results are obtain by setting the algorithm with 10 collocation points per sub-domain, 10 neurons, and sub-domain length of 0.1.
\begin{table}[h!]
\centering
\begin{tabular}{lccccc}
\hline
\multicolumn{1}{c}{} &  \multicolumn{5}{c}{\textbf{MAE}}  \\ \hline
\multicolumn{1}{c}{\textbf{data points}}  & 200 &500 & 1000 & 2000 &  5000  \\ \hline 
$f(x,y,z)$ & 6.50e-00 & 1.58e-01 & 8.57e-03 & 9.39e-04 & 7.98e-04 \\
\hline
\hline
\end{tabular}
\caption{Lorenz system: Gray-Box X-TFC performance in terms of MAE, by varying the number of data points, by using 10 neurons,1020 collocation points per sub-domain, and subdomain length of 0.1.}\label{tab:GBXTFC_lorenz_fun_10_10_0.1}
\end{table}


\subsection{6D Hyperchaotic System}

\begin{figure}[h!]
    \centering
    \includegraphics[width=\linewidth, trim=120 0 110 0, clip]{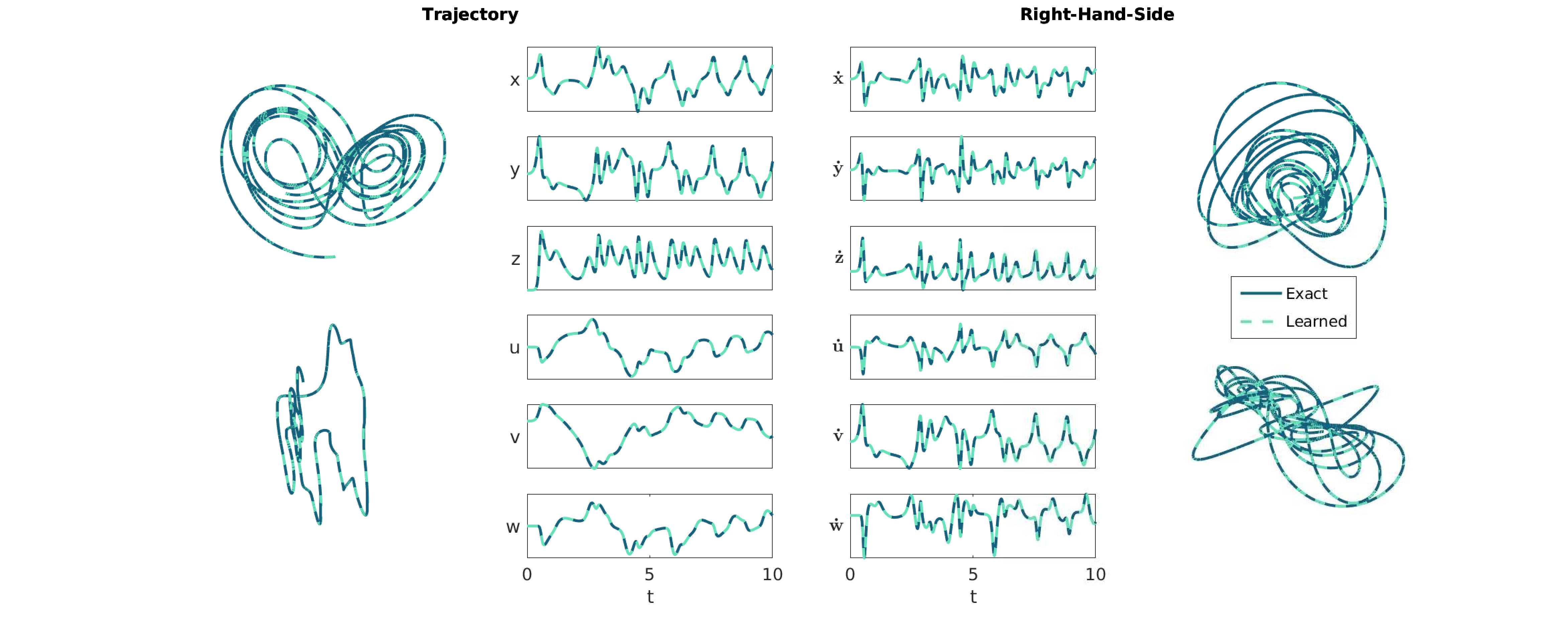}
    \caption{6D Hyperchaotic system: Black-Box X-TFC learned dynamics (left) and RHS (right) vs. exact solutions.}
    \label{fig:bbxtfc_6d}
\end{figure}

The second test case is a 6D hyperchaotic system, modeled by the following set of ODEs
\begin{equation}\label{eq:6D_hyper}
    \begin{cases}
\dot{x} = a(y - x) + u \\
\dot{y} = cx - y -xz - v \\
\dot{z} =  xy - bz  \\
\dot{u} = du - yz \\
\dot{v} = ry \\
\dot{w} = -ew + zu 
    \end{cases} 
    \qquad \qquad \text{s.t.} \qquad \qquad 
    \begin{cases}
        x(0) = 0.1 \\
        y(0) = 0.1  \\
        z(0) = 0.1 \\
        u(0) = 0.1 \\
        v(0) = 0.1 \\
        w(0) = 0.1,
    \end{cases}
\end{equation}
where the parameters are selected as $a=10$, $b=2.6667$, $c=28$, $d=-1$, $e=10$, and $r=3$, for which the system exhibits chaotic behaviour \cite{yi2018dynamical}. The objective of this test case is to prove the capability of our framework by discovering higher dimensional dynamical systems and testing the capability of PySR in distilling the best-fitting expressions from a wider pool of input state variables. The data used for training the NN is generated with 4\textit{-th} order Runge-Kutta method for a time domain $[0,10]$, with time-steps from $\Delta t = 0.05$ to $\Delta t = 0.0002$. \\
The BBX-TFC regression and extraction are performed with 20 neurons, 20 collocation points per sub-domain, and a sub-domain length of 0.1, whose results are reported in Table \ref{tab:bbxtfc_6d}, and qualitative plots are shown in Figure \ref{fig:bbxtfc_6d}. From the table, we can see how increasing the number of data points leads to a decrease in the MAE, going from MAEs in the order of $10^{-5}$ to $10^{-13}$ for the learned dynamics, and from $10^0$ to $10^{-9}$ for the extracted rates of change.

\begin{table}[h!]
\centering
\begin{tabular}{lcccccc}
\hline
\multicolumn{1}{c}{} &  \multicolumn{6}{c}{\textbf{MAE}}  \\ \hline
\multicolumn{1}{c}{\textbf{data points}}  & \multicolumn{1}{c}{\boldsymbol{$x$}}  & \multicolumn{1}{c}{\boldsymbol{$y$}}  & \multicolumn{1}{c}{\boldsymbol{$z$}} & \multicolumn{1}{c}{\boldsymbol{$u$}} &  \multicolumn{1}{c}{\boldsymbol{$v$}} & \multicolumn{1}{c}{\boldsymbol{$w$}}      \\ \hline \hline
200 &  5.22e-06 & 1.12e-05 & 1.40e-05 & 2.34e-05 & 1.56e-06 & 5.63e-05 \\ 
500  &  1.70e-06 & 5.89e-06 & 5.38e-06 & 1.09e-05 & 4.76e-07 & 2.48e-05 \\
1000 &  6.55e-07 & 1.93e-06 & 1.82e-06 & 3.86e-06 & 1.86e-07 & 7.70e-06 \\ 
2000  & 2.00e-11 & 5.42e-11 & 1.19e-11 & 3.92e-11 & 6.12e-12 & 1.04e-10 \\ 
5000  &  1.30e-09 & 3.71e-09 & 3.77e-09 & 7.36e-09 & 4.20e-10 & 1.54e-08 \\ 
10000   & 4.96e-12 & 1.75e-11 & 1.64e-11 & 3.52e-11 & 1.51e-12 & 7.12e-11 \\ 
20000 & 3.75e-12 & 1.07e-11 & 1.06e-11 & 2.14e-11 & 1.11e-12 & 4.23e-11 \\ 
50000  &  4.82e-13 & 2.31e-12 & 3.32e-12 & 1.14e-11 & 1.53e-13 & 2.05e-11 \\ 
\hline
\multicolumn{1}{c}{\textbf{}}   & \multicolumn{1}{c}{\boldsymbol{$\dot x$}}  & \multicolumn{1}{c}{\boldsymbol{$\dot y$}}  & \multicolumn{1}{c}{\boldsymbol{$\dot z$}}    & \multicolumn{1}{c}{\boldsymbol{$\dot u$}} &  \multicolumn{1}{c}{\boldsymbol{$\dot v$}} & \multicolumn{1}{c}{\boldsymbol{$\dot w$}}    \\ \hline \hline
200 &  2.76e-01 & 8.81e-01 & 8.83e-01 & 1.90e-00 & 7.67e-02 & 3.80e-00 \\ 
500  &  1.76e-02 & 5.48e-02 & 5.10e-02 & 1.06e-01 & 5.44e-03 & 2.23e-01 \\  
1000 &   5.58e-03 & 1.30e-02 & 1.45e-02 & 2.74e-02 & 1.68e-03 & 6.10e-02 \\ 
2000  &  1.49e-05 & 2.94e-05 & 2.73e-05 & 4.78e-05 & 4.64e-06 & 1.09e-04 \\ 
5000  & 4.64e-06 & 1.37e-05 & 1.27e-05 & 2.70e-05 & 1.21e-06 & 5.43e-05 \\ 
10000 &  1.10e-07 & 3.59e-07 & 3.79e-07 & 8.08e-07 & 3.10e-08 & 1.42e-06 \\ 
20000 & 2.86e-08 & 6.95e-08 & 7.63e-08 & 1.40e-07 & 8.72e-09 & 3.28e-07 \\ 
50000 &  4.45e-09 & 2.23e-08 & 3.09e-08 & 1.06e-07 & 1.41e-09 & 1.92e-07 \\ 
\hline
\hline
\end{tabular}
\caption{6D Hyperchaotic system: Black-Box X-TFC performance for learned dynamics and learned RHS in terms of MAE, by varying the number of data points.}\label{tab:bbxtfc_6d}
\end{table}

\begin{figure}[h!]
    \centering
    \begin{subfigure}{\textwidth}
        \centering
        
    \end{subfigure}
    \begin{subfigure}{\textwidth}
        \centering
        \includegraphics[width=\linewidth, trim=120 0 0 0, clip]{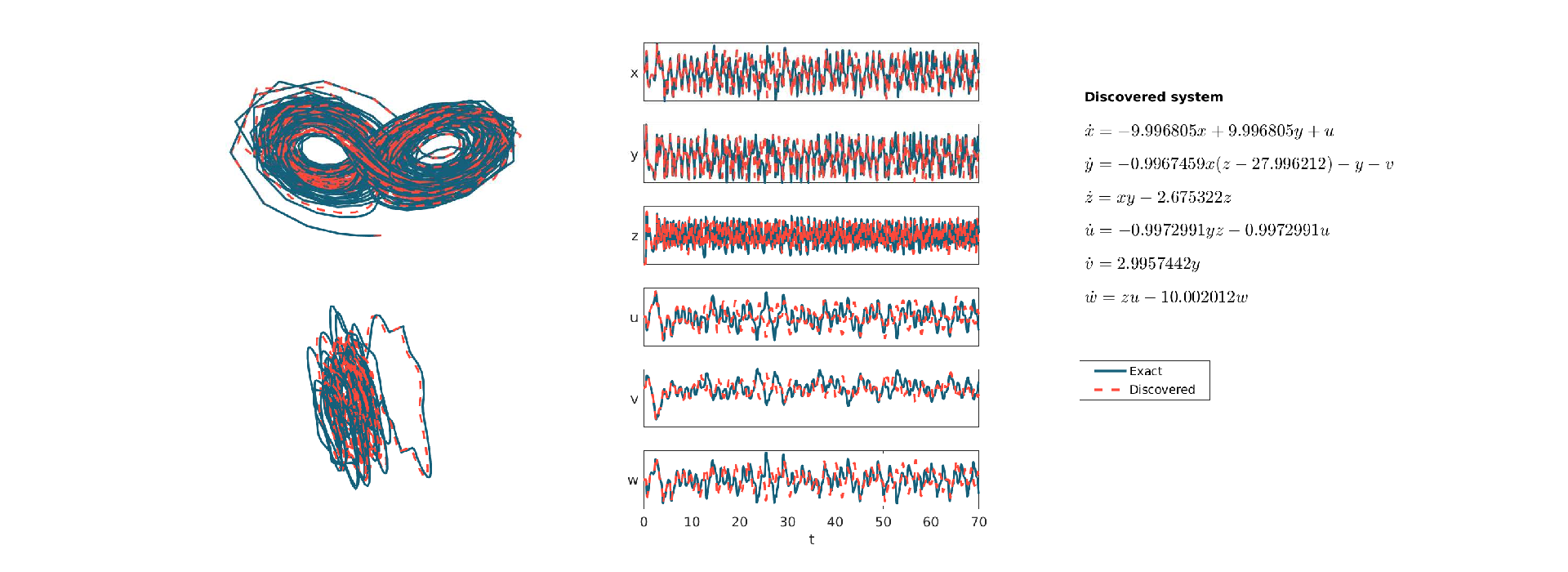}
        \caption{PySR discovered dynamics vs. exact dynamics using Black-Box X-TFC for 200 points.}
        \label{fig:sr_6d_hyper_200}
        \includegraphics[width=\linewidth, trim=120 0 0 0, clip]{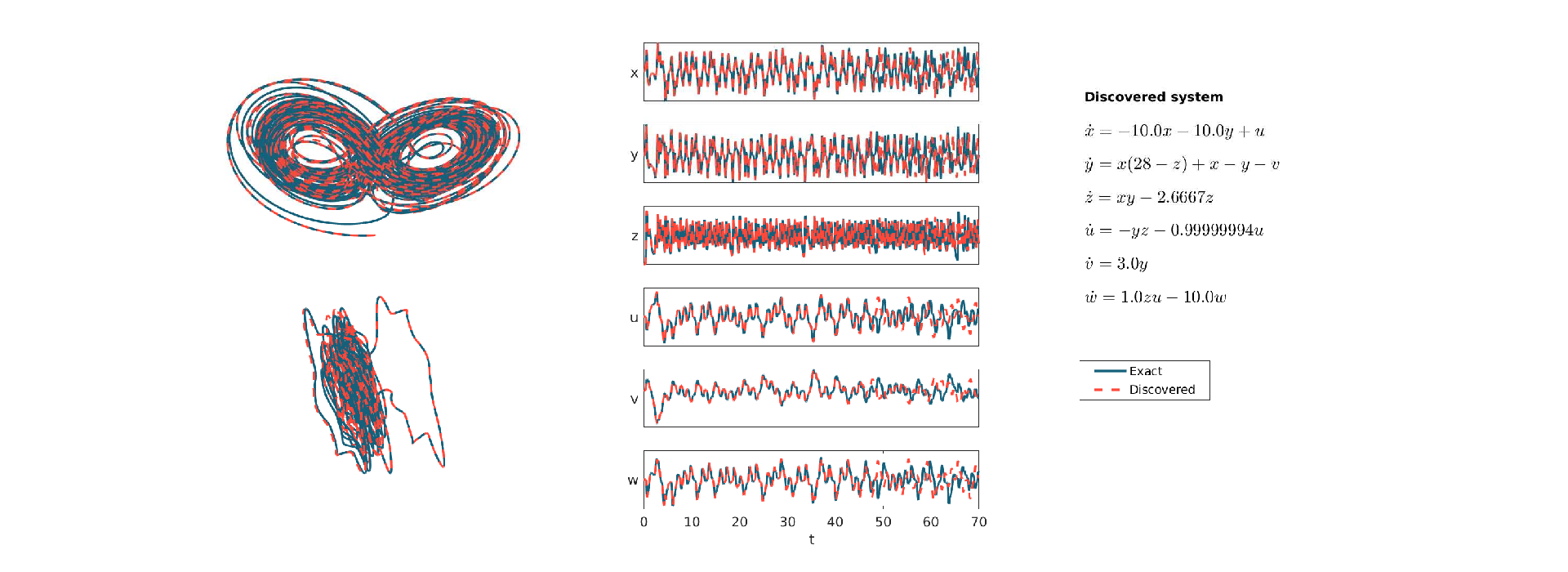}
        \caption{PySR discovered dynamics vs. exact dynamics using Black-Box X-TFC for 5000 points.}
        \label{fig:sr_6d_hyper_5000}
        \includegraphics[width=\linewidth, trim=120 0 0 0, clip]{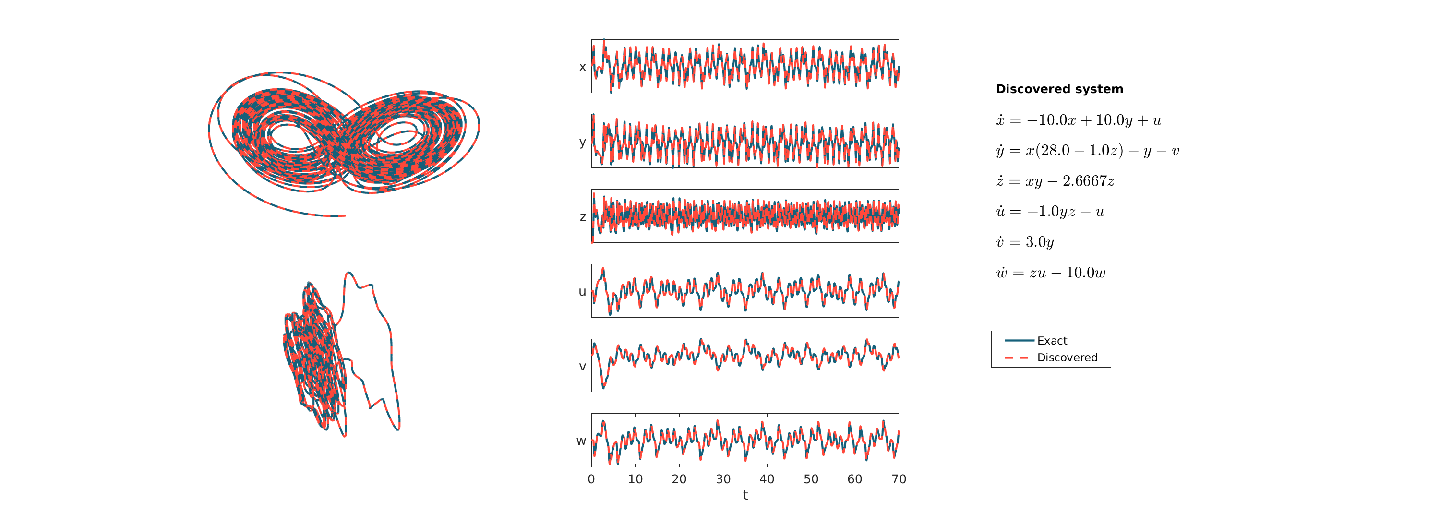}
        \caption{PySR discovered dynamics vs. exact dynamics using Black-Box X-TFC for 10000 points.}
        \label{fig:sr_6d_hyper_10000}
    \end{subfigure}
    \caption{6D Hyperchaotic system: PySR discovered dynamics vs. exact dynamics. The discovered RHS terms of the differential equations are shown on the right.}
    \label{fig:sr_6d}
\end{figure}

The BBX-TFC outputs are used in the PySR algorithm that will find the mathematical expressions made by linear and nonlinear combinations of $x,y,z,u,v$, and $w$ that best fit the functions $\dot x, \dot y, \dot z, \dot u, \dot v$, and $\dot w$. The declared library is made by the binary operators $+$, $-$, and $*$, and population and iteration number set equal to 30.\\
The extracted RHS are then fed into the PySR algorithm to distill the mathematical expressions that best fit the learned dynamics of the system. In Figure \ref{fig:sr_6d}(right), we present the discovered systems of differential equations. Also for this test case, the accuracy of the mathematical expressions and the precision of the discovered parameters depends on the accuracy of the input functions, which depends on the accuracy of the observation data used for the regression. This can be seen in Figure \ref{fig:sr_6d}, where the PySR performance is reported by using 200 (\ref{fig:sr_6d_hyper_200}), 5000 (\ref{fig:sr_6d_hyper_5000}), and 10000 data points (\ref{fig:sr_6d_hyper_10000}). For all cases, the exact equations are discovered, with differences in the precision of the parameters' values, which will obviously lead to a separation between the predicted and the exact trajectory at a certain point in time. In particular, for the case with 200 data points, we are able to predict the dynamics up to about 12 seconds, and further up to about 48 seconds with 5000 data points. When we have enough data points, from 10000 points onwards, the PySR algorithm is able to precisely identify the exact value of the parameters of the differential equations.

\subsection{Non-autonomous Sprott system}

The third selected test case is a non-autonomous, periodically forced chaotic system based on the Sprott C system \cite{sprott1994some}, represented by the following mathematical model:
\begin{equation}\label{eq:sprott}
    \begin{cases}
        \dot{x} = a \sin(y)z + \beta \sin(2\pi \omega t + \Phi_0) \\
        \dot{y} = b (\sin(x) - \sin(y)) \\
        \dot{z} = c - d \sin^2(x)
    \end{cases}
    \qquad \qquad \text{s.t.} \qquad \qquad 
    \begin{cases}
        x(0) = 0.01 + 2 \pi \\
        y(0) = 0.1 + 2\pi  \\
        z(0) = 0.1,
    \end{cases}
\end{equation}
where $\beta \sin(2\pi \omega t + \Phi_0)$ represents the periodic external excitation, in which $\beta$ and $\omega$ are the amplitude and the frequency, respectively, and $\Phi_0 = 0$ is the initial phase. The parameters are selected as $a=3$, $b=1$, $c=0.333$, $d=3$, $\beta=1$ and $\omega=1$, for which the system exhibits chaotic behaviour \cite{wang2020novel}. The data used for training the NN is generated with the 4\textit{-th} order Runge-Kutta method for a time domain $[0,10]$, with time-steps from $\Delta t = 0.1$ to $\Delta t = 0.0005$. 

The BBX-TFC regression and extraction are performed with 10 neurons, 10 collocation points per sub-domain, and a sub-domain length of 0.1, whose results are reported in Table \ref{tab:bbxtfc_sprott}, and qualitative plots are shown in Figure \ref{fig:bbxtfc_6d}. From the table, we can see how increasing the number of data points leads to a decrease of the MAE of the extracted rates of change, for the datasets in the range [100, 500] data points, going from the order of $10^{-3}$ to $10^{-6}$, while it stays between $10^{-6}$ to $10^{-9}$ for richer datasets.

\begin{figure}[h!]
    \centering
    \includegraphics[width=\linewidth, trim=120 0 120 0, clip]{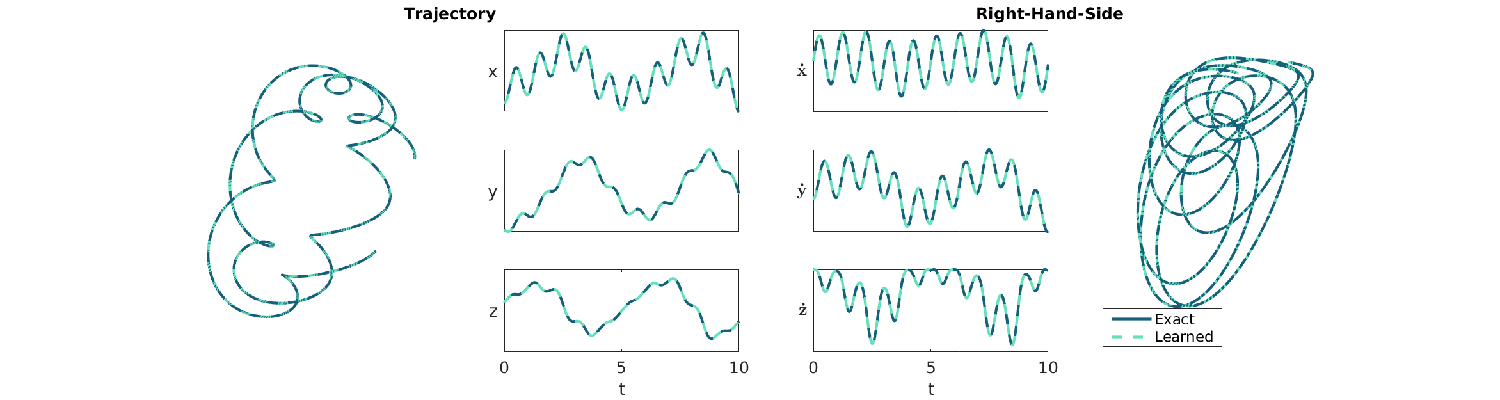}
    \caption{Non-autonomous Sprott system: Black-Box X-TFC learned dynamics (left) and RHS (right) vs. exact solutions.}
    \label{fig:bbxtfc_sprott}
\end{figure}

\begin{table}[h!]
\centering
\begin{tabular}{lccc|ccc}
\hline
\multicolumn{1}{c}{} &  \multicolumn{6}{c}{\textbf{MAE}}  \\ \hline
\multicolumn{1}{c}{\textbf{data points}}  & \multicolumn{1}{c}{\boldsymbol{$x$}}  & \multicolumn{1}{c}{\boldsymbol{$y$}}  & \multicolumn{1}{c|}{\boldsymbol{$z$}}  & \multicolumn{1}{c}{\boldsymbol{$\dot x$}} &  \multicolumn{1}{c}{\boldsymbol{$\dot y$}}   &  \multicolumn{1}{c}{\boldsymbol{$\dot z$}}    \\ \hline \hline
100 & 0.00 & 0.00 & 8.54e-18 & 1.37e-03 & 1.95e-04 & 6.06e-04 \\ 
200 &  1.54e-15 & 2.41e-16 & 6.22e-16 & 1.67e-04 & 2.22e-05 & 6.02e-05 \\ 
500 &   1.77e-17 & 0.00 & 2.45e-17 & 1.55e-05 & 2.14e-06 & 5.98e-06 \\ 
1000 &   0.00 & 0.00 & 4.65e-18 & 3.32e-07 & 4.89e-08 & 9.86e-08 \\ 
2000 &   0.00 & 0.00 & 2.14e-18 & 2.92e-08 & 4.62e-09 & 1.02e-08 \\
5000 &   0.00 & 0.00 & 3.37e-18 & 1.65e-06 & 2.50e-07 & 4.50e-07 \\ 
10000 &  0.00 & 0.00 & 4.47e-18 & 2.31e-07 & 3.58e-08 & 8.22e-08 \\
20000 &   0.00 & 0.00 & 1.73e-18 & 3.95e-07 & 6.79e-08 & 1.31e-07 \\ 
\hline
\hline
\end{tabular}
\caption{Non-autonomous Sprott system: Black-Box X-TFC performance for learned dynamics and learned right-hand-sides in terms of MAE, by varying the number of data points, by using 10 neurons, 10 collocation points per sub-domain, and subdomain length of 0.1.}\label{tab:bbxtfc_sprott}
\end{table}

\begin{figure}[h!]
    \centering
    \begin{subfigure}{\textwidth}
        \centering
        
    \end{subfigure}
    \begin{subfigure}{\textwidth}
        \centering
        \includegraphics[width=\linewidth, trim=120 0 0 0, clip]{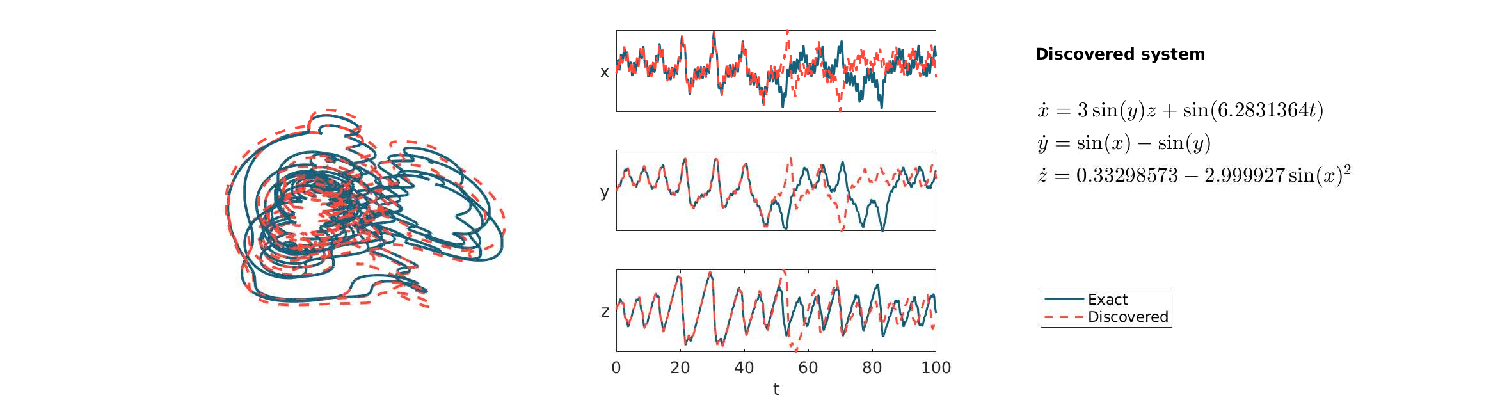}
        \caption{PySR discovered dynamics vs. exact dynamics using Black-Box X-TFC for 200 points.}
        \label{fig:sr_sprott_200}
        \includegraphics[width=\linewidth, trim=120 0 0 0, clip]{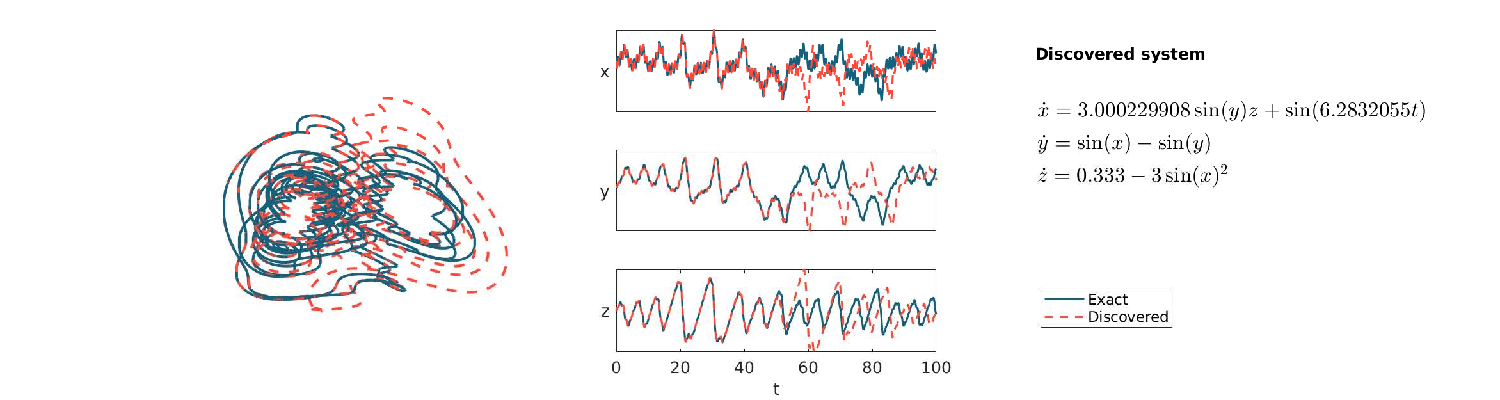}
        \caption{PySR discovered dynamics vs. exact dynamics using Black-Box X-TFC for 2000 points.}
        \label{fig:sr_sprott_2000}
        \includegraphics[width=\linewidth, trim=120 0 0 0, clip]{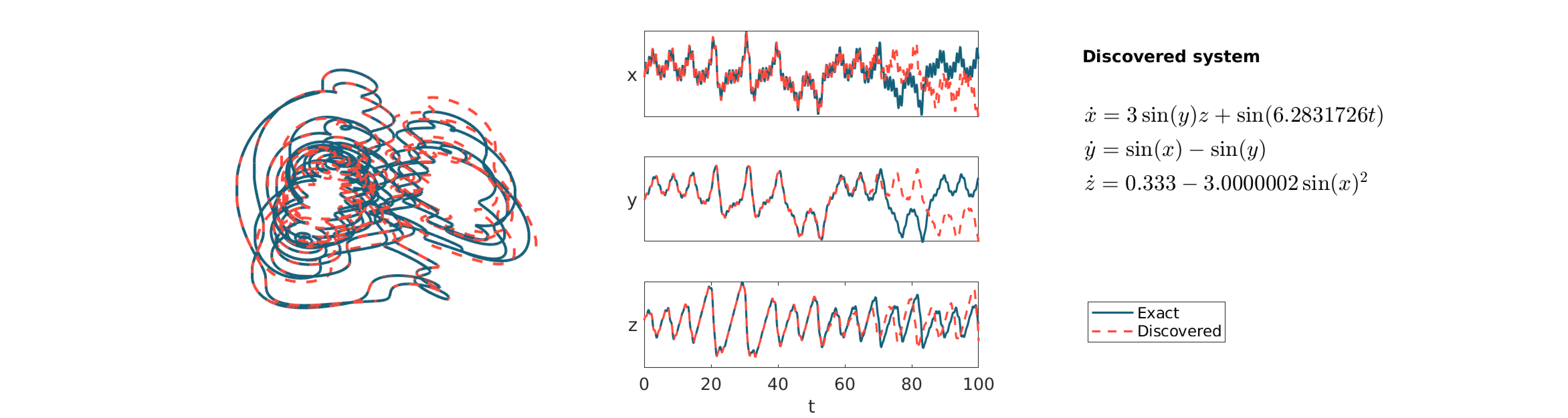}
        \caption{PySR discovered dynamics vs. exact dynamics using Black-Box X-TFC for 20000 points.}
        \label{fig:sr_sprott_20000}
    \end{subfigure}
    \caption{Non-autonomous Sprott system: PySR discovered dynamics vs. exact dynamics. The discovered RHS terms of the differential equations are shown on the right.}
    \label{fig:sr_sprott}
\end{figure}

The BBX-TFC outputs are used in the PySR algorithm that will find the mathematical expressions made by linear and nonlinear combinations of $x,y$, and $z$ that best fit the functions $\dot x, \dot y$, and $, \dot z$. The PySR hyperparameters used for this simulation are a population of 50, and different iteration numbers per equation, such as 500, 30, and 50, for the first, second, and third DE, respectively. For this test case, we also need to declare $"sin"$, and $"square"$ in the unary operators dictionary.\\
The extracted RHS are then fed into the PySR algorithm to distill the mathematical expressions that best fit the learned dynamics of the system. In Figure \ref{fig:sr_sprott}, the discovered systems of differential equations are displayed on the right, producing the predicted dynamics. For this test case, we can see that the accuracy of the mathematical expressions and the precision of the discovered parameters do not change drastically for richer datasets. This can be seen in Figure \ref{fig:sr_sprott}, where the PySR performances are reported by using 200 (\ref{fig:sr_sprott_200}), 2000 (\ref{fig:sr_sprott_2000}), and 20000 data points (\ref{fig:sr_sprott_20000}). For all cases, the exact equations are discovered. In particular, for the case with 200 and 2000 data points, we are able to predict the Sprott trajectory up to about 50 seconds, and further up to about 70 seconds with 20000 data points.

\clearpage

\section{Summary and Discussion}\label{sec:conclusions}

In this paper, we presented an efficient framework based on neural networks and symbolic regression techniques to identify non-linear dynamics of black-box and gray-box models from sparse and noisy time series data, without any (or with partial) prior knowledge of the physics governing the dynamical system's behavior.

The framework consists of two components. The first one is the X-TFC algorithm, which is a minimalistic neural network for forward modeling, data regression, and rate of change and missing term extraction, showing good performance in terms of both accuracy and computational time. The second component is represented by the PySR package, a symbolic regression algorithm that is capable of efficiently distilling mathematical expressions that best fit the RHS of the differential equations modeling the physical phenomenon. The synergy of these two engines resulted in a framework that can accurately discover physical laws from observational data only. 

Its robustness and efficiency were tested by benchmark test cases of autonomous and non-autonomous, chaotic, and hyperchaotic systems. As shown in the Lorenz system example, BBX-TFC can accurately and efficiently extract the RHS with mean absolute errors up to $10^{-9}$, which outperforms a regressor competitor like Gaussian Process Regression. Thanks to its precision, BBX-TFC is a good candidate for extracting the dataset that will feed the PySR algorithm. We demonstrated the robustness of the framework for measurements affected by different levels of noise, being able to make predictions of the chaotic trajectory of the system. Although the test cases in this paper are limited to autonomous and non-autonomous, chaotic, and hyperchaotic systems, our framework can be applied to every time series dataset to discover mathematical expressions that best fit the observations. Hence, this framework of large applicability can provide a model to help forecast the dynamical behavior ahead of time, finding applications in a wide range of real-world scenarios, such as systems biology, financial modeling, orbital mechanics, weather forecasting, and so on. 

Some limitations of this framework remain to be addressed in future studies. Despite the advances in the precision of symbolic regression algorithms in the past few years, there is still a lot of room for improvement, particularly in handling sparse and noisy data, which is very common in real-world scenarios, and adapting to dynamic and evolving systems. As the use of symbolic regression in systems identification becomes more and more ubiquitous, it should become increasingly interesting to test its robustness. This might involve constructing  functions that can ``easily fool" current successful SR algorithms, e.g. functions that feature sharp but smoothly differentiable transitions between different ``symbolically simple" approximations in different portions of a domain.

Additionally, we plan to extend the framework to include differential algebraic equations (DAEs), to investigate the possibility of discovering their related ODEs. Further, it will be worth investigating the application of symbolic regression to model predictive control methods representing the behavior of dynamical systems. The resulting expressions, while potentially complex, would broaden the framework's scope and utility across various scientific disciplines. Future steps for advancing the framework will be using delay embedding theorems to reconstruct the dynamical systems from partial observations of the chaotic attractors and the extension to PDEs and systems of PDEs, to allow scientists from diverse fields of research to benefit from it.

\section*{Acknowledgements}
M.D. and G.E.K. were supported by the Office of Naval Research (ONR) Vannevar Bush grant N00014-22-1-2795. I.G.K. acknowledges partial support from the US Air Force Office of Scientific
Research (AFOSR) FA9550-21-0317 and the US Department of Energy SA22-0052-S001. The codes used for this work can be found, freely available, at the AI-Lorenz GitHub repository \href{https://github.com/mariodeflorio/AI-Lorenz}{https://github.com/mariodeflorio/AI-Lorenz}

\bibliographystyle{ieeetr}
\bibliography{bibliography}

\begin{thebibliography}{10}

\bibitem{rico1994continuous}
R.~Rico-Martinez, J.~Anderson, and I.~Kevrekidis, ``Continuous-time nonlinear signal processing: a neural network based approach for gray box identification,'' in {\em Proceedings of IEEE Workshop on Neural Networks for Signal Processing}, pp.~596--605, IEEE, 1994.

\bibitem{garcia2022machine}
P.~Garc{\'\i}a, ``A machine learning based control of chaotic systems,'' {\em Chaos, Solitons \& Fractals}, vol.~155, p.~111630, 2022.

\bibitem{yuan2023multi}
Q.~Yuan, J.~Zhang, H.~Wang, C.~Gu, and H.~Yang, ``A multi-scale transition matrix approach to chaotic time series,'' {\em Chaos, Solitons \& Fractals}, vol.~172, p.~113589, 2023.

\bibitem{sun2023chaotic}
Y.~Sun, L.~Zhang, and M.~Yao, ``Chaotic time series prediction of nonlinear systems based on various neural network models,'' {\em Chaos, Solitons \& Fractals}, vol.~175, p.~113971, 2023.

\bibitem{clemson2014discerning}
P.~T. Clemson and A.~Stefanovska, ``Discerning non-autonomous dynamics,'' {\em Physics Reports}, vol.~542, no.~4, pp.~297--368, 2014.

\bibitem{hudson1990nonlinear}
J.~Hudson, M.~Kube, R.~Adomaitis, I.~Kevrekidis, A.~Lapedes, and R.~Farber, ``Nonlinear signal processing and system identification: applications to time series from electrochemical reactions,'' {\em Chemical Engineering Science}, vol.~45, no.~8, pp.~2075--2081, 1990.

\bibitem{krischer1993model}
K.~Krischer, R.~Rico-Mart{\'\i}nez, I.~Kevrekidis, H.~Rotermund, G.~Ertl, and J.~Hudson, ``Model identification of a spatiotemporally varying catalytic reaction,'' {\em AIChE Journal}, vol.~39, no.~1, pp.~89--98, 1993.

\bibitem{kevrekidis1994global}
I.~Kevrekidis, R.~Rico-Martinez, R.~Ecke, R.~Farber, and A.~Lapedes, ``{Global bifurcations in Rayleigh-B{\'e}nard convection. Experiments, empirical maps and numerical bifurcation analysis},'' {\em Physica D: Nonlinear Phenomena}, vol.~71, no.~3, pp.~342--362, 1994.

\bibitem{rico1992discrete}
R.~Rico-Martinez, K.~Krischer, I.~Kevrekidis, M.~Kube, and J.~Hudson, ``{Discrete-vs. continuous-time nonlinear signal processing of Cu electrodissolution data},'' {\em Chemical Engineering Communications}, vol.~118, no.~1, pp.~25--48, 1992.

\bibitem{gonzalez1998identification}
R.~Gonz{\'a}lez-Garc{\'\i}a, R.~Rico-Mart{\`\i}nez, and I.~G. Kevrekidis, ``{Identification of distributed parameter systems: A neural net based approach},'' {\em Computers \& chemical engineering}, vol.~22, pp.~S965--S968, 1998.

\bibitem{zhu2023implementation}
A.~Zhu, T.~Bertalan, B.~Zhu, Y.~Tang, and I.~G. Kevrekidis, ``{Implementation and (Inverse Modified) Error Analysis for implicitly-templated ODE-nets},'' {\em arXiv preprint arXiv:2303.17824}, 2023.

\bibitem{cui2023data}
T.~Cui, T.~S. Bertalan, N.~Ndahiro, P.~Khare, M.~Betenbaugh, C.~Maranas, and I.~G. Kevrekidis, ``{Data-driven and Physics Informed Modelling of Chinese Hamster Ovary Cell Bioreactors},'' {\em arXiv preprint arXiv:2305.03257}, 2023.

\bibitem{yin2021augmenting}
Y.~Yin, V.~Le~Guen, J.~Dona, E.~de~B{\'e}zenac, I.~Ayed, N.~Thome, and P.~Gallinari, ``Augmenting physical models with deep networks for complex dynamics forecasting,'' {\em Journal of Statistical Mechanics: Theory and Experiment}, vol.~2021, no.~12, p.~124012, 2021.

\bibitem{malani2023some}
S.~Malani, T.~S. Bertalan, T.~Cui, J.~L. Avalos, M.~Betenbaugh, and I.~G. Kevrekidis, ``{Some of the variables, some of the parameters, some of the times, with some physics known: Identification with partial information},'' {\em arXiv preprint arXiv:2304.14214}, 2023.

\bibitem{li2021grey}
Y.~Li, Z.~O'Neill, L.~Zhang, J.~Chen, P.~Im, and J.~DeGraw, ``{Grey-box modeling and application for building energy simulations-A critical review},'' {\em Renewable and Sustainable Energy Reviews}, vol.~146, p.~111174, 2021.

\bibitem{thilker2021non}
C.~A. Thilker, P.~Bacher, H.~G. Bergsteinsson, R.~G. Junker, D.~Cali, and H.~Madsen, ``Non-linear grey-box modelling for heat dynamics of buildings,'' {\em Energy and Buildings}, vol.~252, p.~111457, 2021.

\bibitem{chen2018neural}
R.~T. Chen, Y.~Rubanova, J.~Bettencourt, and D.~K. Duvenaud, ``Neural ordinary differential equations,'' {\em Advances in neural information processing systems}, vol.~31, 2018.

\bibitem{linot2023stabilized}
A.~J. Linot, J.~W. Burby, Q.~Tang, P.~Balaprakash, M.~D. Graham, and R.~Maulik, ``Stabilized neural ordinary differential equations for long-time forecasting of dynamical systems,'' {\em Journal of Computational Physics}, vol.~474, p.~111838, 2023.

\bibitem{fronk2023interpretable}
C.~Fronk and L.~Petzold, ``Interpretable polynomial neural ordinary differential equations,'' {\em Chaos: An Interdisciplinary Journal of Nonlinear Science}, vol.~33, no.~4, 2023.

\bibitem{goyal2022discovery}
P.~Goyal and P.~Benner, ``{Discovery of nonlinear dynamical systems using a Runge-Kutta inspired dictionary-based sparse regression approach},'' {\em Proceedings of the Royal Society A}, vol.~478, no.~2262, p.~20210883, 2022.

\bibitem{lee2022structure}
K.~Lee, N.~Trask, and P.~Stinis, ``Structure-preserving sparse identification of nonlinear dynamics for data-driven modeling,'' in {\em Mathematical and Scientific Machine Learning}, pp.~65--80, PMLR, 2022.

\bibitem{brunton2016discovering}
S.~L. Brunton, J.~L. Proctor, and J.~N. Kutz, ``Discovering governing equations from data by sparse identification of nonlinear dynamical systems,'' {\em Proceedings of the national academy of sciences}, vol.~113, no.~15, pp.~3932--3937, 2016.

\bibitem{tibshirani1996regression}
R.~Tibshirani, ``Regression shrinkage and selection via the lasso,'' {\em Journal of the Royal Statistical Society: Series B (Methodological)}, vol.~58, no.~1, pp.~267--288, 1996.

\bibitem{donoho2006compressed}
D.~L. Donoho, ``Compressed sensing,'' {\em IEEE Transactions on information theory}, vol.~52, no.~4, pp.~1289--1306, 2006.

\bibitem{champion2019data}
K.~Champion, B.~Lusch, J.~N. Kutz, and S.~L. Brunton, ``Data-driven discovery of coordinates and governing equations,'' {\em Proceedings of the National Academy of Sciences}, vol.~116, no.~45, pp.~22445--22451, 2019.

\bibitem{proctor2014exploiting}
J.~L. Proctor, S.~L. Brunton, B.~W. Brunton, and J.~Kutz, ``Exploiting sparsity and equation-free architectures in complex systems,'' {\em The European Physical Journal Special Topics}, vol.~223, no.~13, pp.~2665--2684, 2014.

\bibitem{bakarji2023discovering}
J.~Bakarji, K.~Champion, J.~Nathan~Kutz, and S.~L. Brunton, ``Discovering governing equations from partial measurements with deep delay autoencoders,'' {\em Proceedings of the Royal Society A}, vol.~479, no.~2276, p.~20230422, 2023.

\bibitem{wei2022sparse}
B.~Wei, ``Sparse dynamical system identification with simultaneous structural parameters and initial condition estimation,'' {\em Chaos, Solitons \& Fractals}, vol.~165, p.~112866, 2022.

\bibitem{udrescu2020ai}
S.-M. Udrescu and M.~Tegmark, ``{AI Feynman: A physics-inspired method for symbolic regression},'' {\em Science Advances}, vol.~6, no.~16, p.~eaay2631, 2020.

\bibitem{cornelio2023combining}
C.~Cornelio, S.~Dash, V.~Austel, T.~R. Josephson, J.~Goncalves, K.~L. Clarkson, N.~Megiddo, B.~El~Khadir, and L.~Horesh, ``{Combining data and theory for derivable scientific discovery with AI-Descartes},'' {\em Nature Communications}, vol.~14, no.~1, p.~1777, 2023.

\bibitem{marra2019constraint}
G.~Marra, F.~Giannini, M.~Diligenti, and M.~Gori, ``Constraint-based visual generation,'' in {\em Artificial Neural Networks and Machine Learning--ICANN 2019: Image Processing: 28th International Conference on Artificial Neural Networks, Munich, Germany, September 17--19, 2019, Proceedings, Part III 28}, pp.~565--577, Springer, 2019.

\bibitem{scott2020lgml}
J.~Scott, M.~Panju, and V.~Ganesh, ``{LGML: logic guided machine learning (student abstract)},'' in {\em Proceedings of the AAAI Conference on Artificial Intelligence}, vol.~34, pp.~13909--13910, 2020.

\bibitem{ashok2021logic}
D.~Ashok, J.~Scott, S.~J. Wetzel, M.~Panju, and V.~Ganesh, ``Logic guided genetic algorithms (student abstract),'' in {\em Proceedings of the AAAI Conference on Artificial Intelligence}, vol.~35, pp.~15753--15754, 2021.

\bibitem{daryakenaria2023ai}
N.~A. Daryakenari, M.~De~Florio, K.~Shukla, and G.~E. Karniadakis, ``{AI-Aristotle: A Physics-Informed framework for Systems Biology Gray-Box Identification},'' {\em arXiv preprint arXiv:2310.01433}, 2023.

\bibitem{schiassi2021extreme}
E.~Schiassi, R.~Furfaro, C.~Leake, M.~De~Florio, H.~Johnston, and D.~Mortari, ``Extreme theory of functional connections: A fast physics-informed neural network method for solving ordinary and partial differential equations,'' {\em Neurocomputing}, vol.~457, pp.~334--356, 2021.

\bibitem{de2022physics}
M.~De~Florio, E.~Schiassi, and R.~Furfaro, ``Physics-informed neural networks and functional interpolation for stiff chemical kinetics,'' {\em Chaos: An Interdisciplinary Journal of Nonlinear Science}, vol.~32, no.~6, 2022.

\bibitem{raissi2019physics}
M.~Raissi, P.~Perdikaris, and G.~E. Karniadakis, ``Physics-informed neural networks: A deep learning framework for solving forward and inverse problems involving nonlinear partial differential equations,'' {\em Journal of Computational physics}, vol.~378, pp.~686--707, 2019.

\bibitem{cranmer2023interpretable}
M.~Cranmer, ``{Interpretable machine learning for science with PySR and SymbolicRegression. jl},'' {\em arXiv preprint arXiv:2305.01582}, 2023.

\bibitem{stephens2015gplearn}
T.~Stephens, ``{gplearn: Genetic programming in python, with a scikitlearn inspired api. [Online]. Available: https://github.com/trevorstephens/gplearn},'' 2015.

\bibitem{boddupalli2023symbolic}
N.~Boddupalli, T.~Matchen, and J.~Moehlis, ``Symbolic regression via neural networks,'' {\em Chaos: An Interdisciplinary Journal of Nonlinear Science}, vol.~33, no.~8, 2023.

\bibitem{mortari2017theory}
D.~Mortari, ``{The theory of connections: Connecting points},'' {\em Mathematics}, vol.~5, no.~4, p.~57, 2017.

\bibitem{leake2020multivariate}
C.~Leake, H.~Johnston, and D.~Mortari, ``The multivariate theory of functional connections: Theory, proofs, and application in partial differential equations,'' {\em Mathematics}, vol.~8, no.~8, p.~1303, 2020.

\bibitem{de2021theory}
M.~De~Florio, E.~Schiassi, A.~D’Ambrosio, D.~Mortari, and R.~Furfaro, ``{Theory of functional connections applied to linear ODEs subject to integral constraints and linear ordinary integro-differential equations},'' {\em Mathematical and Computational Applications}, vol.~26, no.~3, p.~65, 2021.

\bibitem{mai2022theory}
T.~Mai and D.~Mortari, ``Theory of functional connections applied to quadratic and nonlinear programming under equality constraints,'' {\em Journal of Computational and Applied Mathematics}, vol.~406, p.~113912, 2022.

\bibitem{mortari2017least}
D.~Mortari, ``Least-squares solution of linear differential equations,'' {\em Mathematics}, vol.~5, no.~4, p.~48, 2017.

\bibitem{schiassi2022physics}
E.~Schiassi, M.~De~Florio, B.~D. Ganapol, P.~Picca, and R.~Furfaro, ``Physics-informed neural networks for the point kinetics equations for nuclear reactor dynamics,'' {\em Annals of Nuclear Energy}, vol.~167, p.~108833, 2022.

\bibitem{koza1994genetic}
J.~R. Koza, ``Genetic programming as a means for programming computers by natural selection,'' {\em Statistics and computing}, vol.~4, pp.~87--112, 1994.

\bibitem{lorenz1963deterministic}
E.~N. Lorenz, ``Deterministic nonperiodic flow,'' {\em Journal of atmospheric sciences}, vol.~20, no.~2, pp.~130--141, 1963.

\bibitem{sun2023pisl}
F.~Sun, Y.~Liu, Q.~Wang, and H.~Sun, ``{PiSL: Physics-informed Spline Learning for data-driven identification of nonlinear dynamical systems},'' {\em Mechanical Systems and Signal Processing}, vol.~191, p.~110165, 2023.

\bibitem{raissi2018multistep}
M.~Raissi, P.~Perdikaris, and G.~E. Karniadakis, ``Multistep neural networks for data-driven discovery of nonlinear dynamical systems,'' {\em arXiv preprint arXiv:1801.01236}, 2018.

\bibitem{williams1995gaussian}
C.~Williams and C.~Rasmussen, ``Gaussian processes for regression,'' {\em Advances in neural information processing systems}, vol.~8, 1995.

\bibitem{rasmussen2003gaussian}
C.~E. Rasmussen, ``Gaussian processes in machine learning,'' in {\em Summer school on machine learning}, pp.~63--71, Springer, 2003.

\bibitem{seeger2004gaussian}
M.~Seeger, ``Gaussian processes for machine learning,'' {\em International journal of neural systems}, vol.~14, no.~02, pp.~69--106, 2004.

\bibitem{williams2006gaussian}
C.~K. Williams and C.~E. Rasmussen, {\em Gaussian processes for machine learning}, vol.~2.
\newblock MIT press Cambridge, MA, 2006.

\bibitem{galioto2020bayesian}
N.~Galioto and A.~A. Gorodetsky, ``{Bayesian system ID: optimal management of parameter, model, and measurement uncertainty},'' {\em Nonlinear Dynamics}, vol.~102, no.~1, pp.~241--267, 2020.

\bibitem{yi2018dynamical}
L.~Yi, W.~Xiao, W.~Yu, and B.~Wang, ``Dynamical analysis, circuit implementation and deep belief network control of new six-dimensional hyperchaotic system,'' {\em Journal of Algorithms \& Computational Technology}, vol.~12, no.~4, pp.~361--375, 2018.

\bibitem{sprott1994some}
J.~C. Sprott, ``Some simple chaotic flows,'' {\em Physical review E}, vol.~50, no.~2, p.~R647, 1994.

\bibitem{wang2020novel}
M.~Wang, J.~Li, X.~Zhang, H.~H.-C. Iu, T.~Fernando, Z.~Li, and Y.~Zeng, ``{A novel non-autonomous chaotic system with infinite 2-D lattice of attractors and bursting oscillations},'' {\em IEEE Transactions on Circuits and Systems II: Express Briefs}, vol.~68, no.~3, pp.~1023--1027, 2020.

\end{thebibliography}

\end{document}